**Corporate Environmental Management Accounting Practicing and Reporting   in Bangladesh**

Professor Dr. Md. Nazrul Islam
Dr Syed Khaled Rahman,
Department of Business Administration
Shahjalal University of Science and Technology, Sylhet.

This project was submitted to the Shahjalal University of Science and Technology (SUST) Research Centre for financial support.

# ABSTRACT

In the management of environment the Environmental Management Accounting (EMA) is essential for corporate or companies because corporate sectors are the main parties of environmental humiliation as they are existed in the environment and for protecting environment a branch of accounting is emerged which is called environmental management accounting. The objective of the study is to develop a compliance framework for EMA and appraise the ER practices in selected industries in Bangladesh. In conducting the study, 50 environmental sensitive industries were selected from DSE. A compliance checklist was developed on 75 aspects of EMA and ER under 13 groups. In developing the compliance index binary method is used i.e. 1= if ER practices; 0= if not practices. Further the level of EMR/ER practices have been evaluated in terms of selected independent variables of the company viz. total assets, total sales, return on equity and size of board. The study found that the environmental management accounting in the manufacturing companies is in poor level. The maximum compliance is 67% and the lowest is 20%. The TA, TS BS and SP have been considered to find out the explanatory variables. In most of the cases board size does not play significant role in the practice of EMA in the sampled firms.

**CHAPTER ONE**

**INTRODUCTORY ASPECTS OF THE PROJECT**

**1.1. Background of the Study**: The aim of first  global conference  on environment of United Nations held in 1972, in Stockholm was to  protect and improve the human environment and to remedy and prevent its  impact and the outcome was  a draft on "Universal Declaration on the Protection and Preservation of the Human Environment"   (UNCED,1992). After 20 years the second conference was  held  in  Rio Di Janeiro    on environment and development -called Earth Summit 1992 where 27 principles were adopted by the reaffirming the declaration  of the first  conference  of  1972(  Finance  Division,  2014;  United,  Conference,  Conference,  & Environment, 1992). Lastly  after 20 years the third conference was held in  Rio de Janeiro in 2012   on  sustainable  development-  called   Rio+20  where a    document  of  commitment containing    clear  and  practical  measures  for  implementing  sustainable  development  viz.  economic growth, social improvement and environmental protection align  to the  goals  of Earth Summit in 1992(Francis & Si, 2015; Finance Division, 2014).   A set of Sustainable Development Goals (SDGs) were build upon the Millennium Development Goals . Among the 17 SDGs, SDG6-clean water and sanitation, SDG7-sustainable energy for all, SDG8-decent work and economic development, SDG9- innovation and infrastructure and SDG12-sustainable consumption and protection  are directly related to environmental management (United et al., 1992). In  the  management of environment the  Environmental Management   Accounting  is essential   for  corporate  or  companies   because  corporate  sectors  are  the  main  parties  of environmental  humiliation  as  they  are  existed  in  the  environment  and  for  protecting environment, accounting is emerged  which is called Environmental Management Accounting (EMA).

 The International Federation of Accountants(Johnson, 2004),  defined defines Environmental Management  Accounting (EMA) as -

"the management of environmental and economic performance through the development and implementation of appropriate environment-related accounting systems and practices. While this may include reporting and auditing in some companies, environmental management accounting typically involves life-cycle costing, full-cost accounting, benefits assessment, and strategic planning for environmental management."

Environmental Reporting (ER)   is the output of the  implementation of EMA. Environmental reporting practices of the industrial organizations  are crucial issues for  sustainable development as   many industrial   activities gradually increase environmental hazards.   Many national and international organizations have been working for the development of specific conceptual and regulatory framework such as UNEP and  UNCTD. The most  influential and pioneering effort on environmental reporting is Global Reporting Initiative (GRI)   and ISO 14001:2015. An increasing number of countries are   imposing   requirements on companies for reporting environmental performance. Denmark is the first country to adopt mandatory legislation on public environmental reporting. In Netherlands, new legislation on mandatory environmental reporting has been adopted. Both Danish and Dutch regulations require reporting to the authorities and to the public. In Norway, the new Accounting Act requires that all companies include environmental information in the annual report from 1999 onwards. In Sweden, similar legislation has been adopted for mandatory environmental disclosure in annual report. In U.S.A. the companies are required to submit data on emission of specific toxic chemicals to the Environmental Protection Agency under the Toxic Release Inventory .In Canada, The Securities Commission requires public companies to report the current and future financial or operational effects on environmental protection requirements in an Annual Information Form. GoB enacted laws regarding environment viz. Bangladesh Environment Conservation Act, 1995. The  legal framework for accounting and reporting in  Bangladesh is primarily governed Bangladesh Accounting Standard and Bangladesh Financial Reporting Standard ,Securities and Exchange Commission Rules 1987 and the Income Tax Ordinance-1984. These laws do not prescribe any mandatory environmental  accounting disclosure by the companies. EMA will help in this case and this  is the motivation behind the study.

**1.2.Statement of Problem  of the study**:  Bangladesh is one of the most densely populated countries in the world in which approximately 29 percent of the population lives below the poverty line (MOF, Bangladesh Economic Review-2015). The contribution of industry sector to GDP is progressively increasing in Bangladesh. According to the BBS the contribution of the broad industry sector to GDP has been estimated at 31.54 percent in 2015-16 which was 30.42 percent in 2014-15 highest in GDP. In 2014-15 the contribution of manufacturing sector in GDP was 20.16 percent which increased to 21.01 percent in 2015-16. To provide accommodation for the increasing manpower and reduce the poverty rate, manpower intensive industrialization programmers have been emphasized. That is why corporate sectors of Bangladesh received serious attention in all the Five Year Plans(There are seven five year plans) of the country to accelerate economic growth, increase investment, earning foreign exchange, create employment and reduce poverty. The Perspective Plan of Vision 2010-2021 and Seven Five Year Plan (2016-2021) have also recognized the importance of corporate sector as vehicle for creating productive high income jobs and development (Nath,2012). As a result, the corporate sector grew at a rate of five percent between 1972 and 1992 (Bhattacharya et al., 1995). The growth of industrial activities in Bangladesh has a positive development dynamic in macro-economic terms. The contribution of manufacturing sector to GDP has been increased from 10% in 1970 to 17% in 2010 and 21% in 2015-16. Moreover, employment share of the sector was 11.90% of total employment in 2011 (Nath,2012).

The growth of corporate sector has simultaneously accelerated severe environmental hazards in Bangladesh (SEHD,1998).Many industries do not operate effluent treatment plants (ETP). Garments, textile and dying sectors have been developed without proper attention to their environmental consequences. Others polluting sectors are - tanneries, chemical and pharmaceutical industries and ship-breaking yards. According to Ahmed (2012) Bangladesh is to pay huge environmental costs for its economic development. Because the air, water and noise pollution which are derived from industrial activities threatening human health, ecosystems and overall economic growth of the country. The Department of Environment (DoE) has listed 1,176 factories that cause pollution throughout the country. Industrial growth also creates a range of problems. In consequence rapid and largely unregulated industrial development, many aquatic eco-systems are now under threat and with them the livelihood systems of local people. According to the survey of Society for Environment and Human Development (SEHD) that treatment of industrial wastes was considered a low priority and that due to the absence of strong preventative measures and lack of awareness, the practice of discharging untreated industrial waste into water bodies was almost universal. As a result, acute environmental pollution due to manufacturing activities is now threatening the sustainability of the resources and increasingly impacting on the public health (BSEHD,2001;Alam, 2002). So it is the demand of time to incorporate environmental responsiveness among the organizations (corporate) in Bangladesh. In this regard, practicing and reporting of EMA can play a vital role. But there is no mandatory standard or code in prevailing accounting and reporting system that can consider the environment unfriendly activities of the corporate entities. Moreover, the existing environmental laws and other corporate related laws do not prescribe adequate environmental disclosure to be made by the listed companies in Bangladesh. Avoidance of environmental information in accounting system can create a gap in financial reporting of companies. lf vital environmental issues are not disclosed, the financial statements cannot be considered as true and fair view of affairs . Thus, the traditional reporting system cannot provide the real picture of the organizations' environmental performance, though it is essential for the decision making process of management and stakeholders. On the other hand, company's environmental performance affects its financial health as well as the overall environment of the country. Environmental reporting is a tool for the companies to communicate their environmental performance to the stakeholders. That is why; there is an increasing demand from various stakeholder groups for companies to publicly report information regarding their environmental activities in a global scale (United Nations, 1998).

There has been a growth in the voluntary environmental reporting practices of the corporate organizations worldwide (Kolk, 2003). Some developed countries have initiated mandatory disclosure for corporate organizations. But in Bangladesh, corporate environmental disclosure is still in its nascent stage (Belal, 2001). Bangladeshi companies have been adopting environmental reporting practices voluntarily in recent years . As stated earlier in many countries environmental accounting has become mandatory. But in Bangladesh it is in totally voluntary stage. Further, the companies are reporting voluntary in different forms as there is no guideline or code nationally. The managers of corporate sectors have not clear concept on EMA and also they do not know how to implement EMA. Considering these, the researchers have undertaken the present study to fill the gap of knowledge in this area.

**1.3.Rationality of the study :** The present study is relevant to the following ways:

**1.3.1. Relevant to the protection the environment from business activities:** The environment is changing with the changing of life pattern and economic growth. Many studies found that there is a positive relationship of economic growth and environmental pollution(MIT, 2008)(López, 1994).In Bangladesh economic growth rate has been increasing but with this increasing trend the environmental pollution is also increasing. The growth rate of GDP is more than 7% in 2015-16 which was below 6% in 2010 while the environmental pollution is also increasing(Project, 2015). The business organization is disturbing the natural flow of environmental structure through their thoughtless operations. Therefore, the organizations of business must have the moral commitment towards environment so that the environmental health takes proper consideration and treatment during their operation. The environmental management accounting can help in this case to save the environment from the business operations. And for this, the managers of business organization should have clear concepts on EMA and implementation of EMA. In Bangladesh, there is a huge gap of knowledge on EMA to the mangers of corporate sectors (Project, 2015). The present study will help in filling this gap of knowledge on EMA.

**1.3.2. Relevant to the national sustainability development strategy and long term plans** : The GoB of Bangladesh has prepared the National Sustainable Development Strategy (NSDS) with the help of United Nations Environment Program (UNEP), to meet the alarming environmental challenges and has identified the environment management as one of the five strategic priority area(Planning Commission of the Government of the People's Republic of Bangladesh, 2013) and has articulated a wide variety of actions needed for sustainable development(General Economics Division (GED), Planning Commission, 2015) in which organizational environment management system should be reviewed under the obligation of agenda 21 of SDG(Planning Commission of the Government of the People's Republic of Bangladesh, 2013). To ensure the good governance in environmental sustainability, GoB has taken strategy in the seventh five year plan also (General Economics Division (GED), Planning Commission, 2015). In this case, EMA will help to achieve NSDG which is united with the SDG.

**1.3.3. Relevant to the deveopment of environmental freindly product and services** : To achieve the SDG, we need corporate environmentalism in all functional areas. To this aim, organization should be award with environmentalism and there is a positive relationship between corporate environmentalism and EMA(Larsson & Svensson, 2010).

**1.3.4.     Relevant to the organizatinal sustainability   and efficiency:** For long term sustainability of an organization, it should comply all the environmental rules and regulations that are forced by the monitoring agencies of Bangladesh.  Otherwise, it should not sustain in the long run.  EMA in this case will make easy to confirm the environmental laws, rules and regulations through internal and external environment management. Environmental management accounting   can assist the company in improving efficiency and minimizing waste by understanding environmental costs and their cost drivers.

**1.3.5. Relevant to the Globalization :** As stated earlier, the environmental consiousness around the world is increasing day by day for the sustainability. Many countries alrady haveintroduced mandatory environmental code for orgnizations. The have dveloped the trade  policy after considering the corporate  enviornmentalism and according to this policy, they do not do business with those organization whice do/does not comply the environmentalism according to thir policy. For instance, Bangladesh is one of the RMG exporting country we know, and this sector is under serious pressure from the buyers under corporate  environentalim policy and for this we have lost GSP. Corporate environment management is one the criteria for international business (Larsson & Svensson, 2010).Therefore, to comply the global  busines prsessure Bangladesh should give priority in corporate environment management and EMA in this regard will directly help.

Thus, the project is very  relevant to the global, national and corporate strategic goals from different perspectivs.

**1.4.Objectives of the study**: The   core objective of the study is to  make an appraisal of corporate environment management  accounting system in Bangladesh. The specific objectives are to:

(i) identify  the existing literatures, guidelines codes, and standards on corporate  Environmental Management Accounting ;

(ii) develop a conceptual framework and an Implementation framework of corporate Environmental Management Accounting;

(iii) make an appraisal  on  the practices of corporate Environmental Management Accounting in Bangladesh  in terms of the framework developed;

(iv) identify the   factors responsible  for   implementation of corporate Environmental Management Accounting in Bangladesh; and

(v) recommend   some policy suggestions for the better implementation of corporate Environmental Management Accounting in Bangladesh.

**1.5.Methodology of the study:**

**1.5.1.Population and sample**: The population of the study is all the listed firms of Dhaka Stock Exchange Limited(DSE)  as on Janury 2017. There are 563 listed firms in 22 sectors  in DSE among them Treasury bond 221, mutual fund 35, and corporate bond 2, and debenture  8 are issud by the government and debenture and mutual fund are issued by the exsiting firms. Therefore, in sampling we excluded these 266  from  popuation.Further Bank 30,  Financial institutions 23, Insurance  47; Travel & leisure  4, Miscellaneous 12 are excluded from the study as they are not manufacturing. Therefor net population is (563-266-116)181. Due to budget and time limitation 61 (more than 25%of population ) has been taken for the study while the standard sample size is 123. The population and sample is given in table 1 below table -1: population and sample

| Sl.# | Sector | Total listed Number | Study in population | Sampled firm | Minimum sample size |
|------|--------|---------------------|---------------------|--------------|---------------------|
| 1 | Bank | 30 | Not Included | | |
| 2 | Cement | 7 | 7 | | |
| 3 | Ceramic | 5 | 5 | 3 | 61* |
| 4 | Corporate bond | 2 | Not included | | |
| 5 | Debenture | 8 | Not included | | |
| 6 | Engineering | 33 | 33 | | |
| 7 | Financial institutions | 23 | Not Included | | |
| 8 | Food and allied | 18 | 18 | 10 | |
| 9 | Fuel and power | 18 | 18 | 10 | |
| 10 | Insurance | 47 | Not Included | | |
| 11 | IT sector | 7 | 7 | | |
| 12 | Jute | 3 | 3 | 2 | |
| 13 | Miscellanious | 12 | Not Included | | |
| 14 | Mutual fund | 35 | Not included | | |
| 15 | Paper and print | 2 | 2 | 1 | |
| 16 | Pharmaceuticalsand chemicals | 28 | 28 | 15 | |
| 17 | Service and real esate | 4 | 4 | | |
| 18 | Tannery | 6 | 6 | | |
| 19 | Telecomunication | 2 | 2 | | |
| 20 | Textile | 48 | 48 | 20 | |
| 21 | Travel and leisure | 4 | Not Included | | |
| 22 | Treasury bond | 221 | Not included | | |
| 23 | Total | 563 | 181 | | |

$$* = \frac{\chi 2 NP(1-P)x^2}{\{C2(N-1) + \chi 2\ P(1-P)\}}$$

Where : $\chi 2$ =chi-square ; N= Number of population ; P=probability; C=level of confidence.

**1.5.2. Sources of data and their collection**: The study is based on primary data. The published annual reports of the sampled firms are the main source of secondary data. The content analysis method has been used to collect the secondary data. The content analysis method employed to measure disclosures that has been defined as "a technique of gathering data that consists of codifying qualitative information in anecdotal and literary form, into categories in order to derive quantitative scales of varying levels of complexity" (Abbott and Monsen, 1979, p.504).

**1.5.3. Data Analysis:** In analysis the data following steps have been undertaken:

**1.5.3.1. Constructing the checklist for secondary data:** A checklist has been developed to explore the current practices of environmental management accounting in the sampled firms. The annual reports of five years 2012-2016 have been analyzed for the study. Thus 5*61 = 305 annual reports have been reviewed. In constructing the checklist related theories, literatures, past studies, EMA codes and guidelines have been reviewed and discussion with academicians and professions have been made. The checklist is given in the chapter three.

**1.5.3.2. Scoring the Items in the Checklist**: In order to score the items of the checklist, dichotomous approach (also known as binary approach) has been followed in the present study. Under dichotomous approach, an information item gets a score of one(1) if it is reported and gets a score of zero(0) if it is not reported.

**1.5.3.3. Calculating the Value of Corporate Reporting Index**:

Finally, the Corporate Environmental Management Reporting Index (CEMRI) has been constructed by using the following formula):

$$CEMRI = \frac{\text{Total Number of Items Actually Reported in the Annual Report}}{\text{Maximum Number of Items Expected to be reported in the Annual Report (75)}}$$

**1.5.3.4. Regression Analysis**: The multiple regressions are used to analyze the data. Where CEMRI is the dependent variable and some corporate specific financial and non-financial characteristics are independent variable. The regression model is given below:

To determine the factors affecting on EMA index practice, we use the Multiple Linear Regression Model. The Multiple Linear Regression Model has been postulated as follows:

$$Y = \beta_0 + \beta_1 X_1 + \beta_2 X_2 + \beta_3 X_3 + \beta_4 X_4 + \varepsilon$$

So, $\qquad$ EMA Practices $= \beta_0 + \beta_1 TA + \beta_2 TS + \beta_3 SP + \beta_4 BS + \varepsilon$

Where,

| | | |
|---|---|---|
| Y | = | EMA Practices(CEMRI) |
| TA | = | Total Assets of the company |
| TS | = | Total Sales of the company |
| SP | = | Stock Price |
| BS | = | Board Size |

$\beta_\circ$ is the intercept and $\beta_1$, $\beta_2$, $\beta_3$, $\beta_4$ are regression co-efficient.

# CHAPTER TWO

# LITERATURE REVIEW

Literature review is used to explore the concepts, methods and findings of the past studies done by different scholars in the relevant field so that the researcher can make research in right direction. With this mind available literature are reviewed here on environmental management accounting especially in Bangladesh.

**Karimi et al. (2017)** made a study on "Analysis of Factors Affecting the Adoption and Use of Environmental Management Accounting to Provide a Conceptual Model" to find out the influential factors of environmental management accounting tools to employee level in oil refining and petrochemical companies. They used resistance to change and constraints of gathering and allocating environmental costs and lack of clear standards as internal factors, and competitive environment and culture as external factors. Descriptive survey method was used and its design was quasi-experimental. Structured questionnaire was used and Cronbach's lpha and split-half were used to evaluate the reliability of the questionnaire. 200 questionnaires are distributed and 160 respondents give their opinion. Collected data were analyzed using descriptive(mean, median, mode, standard deviation) and inferential statistical methods i.e. one sample t-test, Pearson correlation, confirmatory factor analysis, path analysis, structured equation modeling, two sample t-test, analysis of variance etc. The authors proposed five hypotheses. They found that from managers and assistants' point of view, the aforementioned factors affected the implementation of EMA tools in the company. They concluded that right people should be appointed who take risks, accept changes procedure fast, and use effective and efficient standards to implement EMA tools in the company. They recommended that future research is needed to identify different patterns of innovation adoption study and develop other factors (Karimi et al., 2017).

**Jamil et al. (2015),** in their study on "Environmental Management Accounting Practices in Small Medium Manufacturing Firms" aimed to find out the practices of EMA in Small Medium Manufacturing Firms. They investigated factors and barriers which influence the practices of EMA. They used questionnaire to collect the data and their focusing area was the Malaysian Small Medium Manufacturing Firms. They found that the firms had a budget allocation for environmental activities and practice physical EMA. Coercion is a dominant factor for practicing EMA (Government pollutions standard, Government regulations, company shareholders, Newspaper and TV, environmental laws etc.) In their findings, some barriers of EMA practice were attitudinal barriers (Low priority of accounting for environmental costs, resistance to change), financial barriers ( resource constraints, efficiency of financial considerations, environmental costs are not considered significant), information barriers (difficulties in collecting or allocating environmental costs, low physical environmental uncertainty), institutional barriers (Lack of institutional pressure, stakeholder power, shareholder power), management barriers (Few incentives to manage environmental costs, Lack of environmental responsibility and accountability etc.). The authors did not focus on the monetary aspects of EMA. So, future studies might be conducted in this field (Jamil et al., 2015).

**Tsui (2014)**, in a study on "A Literature Review on Environmental Management Accounting (EMA) Adoption" aimed to focus sustainability development, EMA, Environmental costs, physical EMA (the flow of water, energy etc.), monetary EMA (the cost of firms' consumption of natural resources, the costs for controlling or preventing environmental damages etc.), EMA adoption etc. in Hong Kong. He found that EMA could help firms to identify cost savings opportunities and to develop more efficient production processes. He also stated that there is still having some problems of the application of EMA at firm level. Studies found that lack of promotion on the use of EMA, lack of collaboration between accountants and environmental management departments were major barriers of EMA adoption. Accountants thought that implementing EMA was costly. Finally, managers avoid the responsibility of EMA adoption barriers for significant environmental costs. Because, in the highly regulated financial standards did not include environmental costs. The study recommended that EMA is better technique to manage the firms' environmental costs and to improve firms' production processes, environmental performance and to achieve sustainable development (Tsui, 2014).

**Klassen & McLaughlin (1996),** in their study on "The Impact of Environmental Management on Firm Performance", aimed to aimed to identify a linkage between strong environmental management to improved perceived future financial performance focusing on financial performance, environmental awards, operations management, and environmental performance of a firm. They told that Environmental Management encompasses all efforts to minimize the negative environmental impact of the firm's products throughout their life cycle. They also told that environmental performance measures how a firm can be succeeded in reducing and minimizing its impact on the environment. The authors develop a theoretical model firstly to show that environmental management is linked interactively to both corporate and functional strategies. Secondly, empirical evidence supported to overall linkage hypothesized by the model. The authors use methodology to find environmental performance by short-term and long-term basis. A standard approach is to employ the market model. The linkage to firm performance is tested empirically to evaluate the data of firm-level environmental and financial performance and using financial event methodology. Cross sectional analysis is used to analyse the data. To conclude, the authors comment that this linkage between Environmental Management and Financial Performance can be used by both researchers and practitioners as one measure of the benefits experienced by industry leaders, and as one criterion against which to measure investment alternatives. Further research is needed in attempting to estimate the difference between the market valuation of a crisis and the costs of the actual clean-up and penalties. This value would capture some of the perceived costs of poor environmental management systems (Klassen & McLaughlin, 1996).

**Ahmed (2012),** in his study titled "Environmental Accounting and Reporting Practices: Significance and Issues: A Case from Bangladeshi Companies" aims to critically evaluate the Environmental Accounting practices in the selected 40 respondents (Chief Accountants and Senior Accountants) of selected companies categorizing into 8 such as Pharmaceuticals, Tannery, Cement, Ceramics, Engineering, Food& Beverage, Textiles, Fuel & Power taking 5 respondents of each. He conducts the study which is based on both the primary and secondary data. He finds that the accountants of the companies strongly need environmental accounting practices and environmental reporting in their Annual Reports (Ahmad, 2012).

**Noah (2017)** conducts a study on "Accounting for the environment: The Accountability of the Nigerian Cement Industry" for analyzing how corporate environmental issues are accounted for the cement industry of Nigeria. He states that environmental polluting has been a concern for among transnational organizations like World Bank, IMF etc. governments, policymakers and society. It is a qualitative research. Case study approach is applied in this study where two companies are selected as case. Semi-structured interview method, documentary evidence and visual methods are used in data collection. The researcher uses semi-structured interview method in order to collect in-depth information from the respondents interviewed, used documentary evidence methods to collect information from annual reports of the companies and from official government documents to know about the commitments and constraints of companies and government to corporate environmental accountability practices. The author found that environment was polluted greatly for the operations of cement industry in Nigeria. He also found that these companies were under pressure from a number of agents for the reduction of environmental pollution. He showed that external institutional factors largely influenced corporate environmental accountability practices in cement industry of Nigeria. That's why cement industry did not show any significant role for the wellbeing of the environment and citizens. The author concluded that he tried to contribute with the existing literature of various researchers in corporate environmental accountability practices. The findings of the study will be helpful for policy implications in Nigerian manufacturing industry, specifically cement industry. Further studies are needed to make corporate environmental accountability practices policy for cement industry from different countries' perspective (Noah, 2017).

**Sendroiu et al. (2006)** made a study on "Environmental Management Accounting (EMA): Reflection of Environmental Factors in the Accounting Processes through the identification of the environmental Costs Attached to Products, Processes and Services" to define the general framework and the conditions to implement EMA in Romanian organization. They attempted to identify the useful information to help minimizing environmental costs and negative impact on environment. For this reason, the paper discusses about EMA by identifying environmental factors in the accounting process through environmental costs identification which are attached to products, processes and services. They also demonstrate impact of environment related activities, to identify cost reduction, to prioritize environmental actions. They defined EMA as the identification, collection, estimation, analysis, internal reporting, and use of materials and energy flow information , the information of environmental costs, and others costs for both environmental and conventional decision making within organization. In this paper, the researchers focused on two elements of sustainable development (environment and economics). They observed that many people overestimate the costs and underestimate the benefits of improving environmental practices. The researchers reached in a conclusion that management accounting techniques can distort and misrepresent environmental issues. They also concluded that these issues can be solved by applying environmental managerial accounting in the organization (Sendroiu et al., 2006).

**Uwuigbe (2012),** in his study on "Web-based Corporate Environmental Reporting in Nigeria: A Study of Listed Companies" aims to identify the fact whether Listed financial and non-financial companies in Nigeria utilize internet for communicating environmental accounting information or not. The author selects 30 firms on the Nigerian Stock Exchange as a sample. He uses content analysis technique to gather data from selected firms' website. T-test statistics is used to show the difference in the level of web-based corporate environmental disclosure between Nigerian financial and non-financial firms. Linear regression method is applied to investigate the relationship between firms' financial performance, size of firms and in the level of corporate environmental disclosure of firms consecutively. The author finds that there is no significant difference in the level of web-based corporate environmental disclosure between financial and non-financial firms in Nigeria. The author point out that online environmental reporting in Nigeria is still in its early stage. If the firms use online communication system to disclose web-based environmental information, it improves the quality of information disclosed to stakeholders (Uwuigbe, 2012).

**Makori & Jagongo (2013)** conducted a research "Environmental Accounting and Firm Profitability: An Empirical Analysis of Selected Firms Listed in Bombay Stock Exchange, India" aimed to identify the significant relationship between environmental accounting and profitability of selected firms Listed in India. They mention about two types of environmental reporting (mandatory disclosure and voluntary disclosure) practices in Indian companies. They gain substantial interest on environmental accounting and reporting from different academic researchers for the past forty years (Rajapakse, 2003; Surman and Keya, 2003; Thompson & Zakarai, 2004; O'Donovan & Gibon, 2002). The studies give the results that there is a relationship between corporate financial performance and corporate social and environmental disclosures. The researchers the literature related to profitability as well as corporate financial performance. They developed a framework including some independent variable to find out firms' Return on Capital Employed (ROCE), Net Profit Margin (NPM), Dividend per Share (DPS), Earnings Per Share (EPS) to measure profitability and its relationship with environmental accounting. To find out this relationship between different variables, the researchers use multiple regression model/analysis to analyse the data. Secondary data were used and collected form annual reports and accounts of 14 randomly selected companies Listed in Bombay Stock Exchange for the year 2007. These 14 companies were drawn from various sectors including automobile, tobacco, chemical, Fertilizer, cement, pharmaceutical, oil and gas exploration and refinery, engineering, Food and personal care products, Glass and ceramics, telecommunications, Cable and Electric Goods, Leather and Tanneries, Synthetic and rayon, textile spinning and textile weaving (Makori & Jagongo, 2013).

**Frost & Wilmshurst (2000)** made a study named "The adoption of Environment-related Management Accounting: An Analysis of Corporate Environmental Sensitivity" aimed to identify environmental sensitivity of the industry of a factor associated with the adoption of environmental-related management accounting and control procedures. They used Deegan and Gordon's methods in the determination of environmentally sensitive industries where the respondents being asked to rank industries on a scale 1-5 (5 being most sensitive). The top 10 specific industries were selected as sample like, uranium mining, chemical, coal, transport, oil/gas explorers, plastics manufacturing, oil/gas producers, gas distributors, paper merchants and timer products as environmentally sensitive industry. The determination was based on indirect method where literature Sources have identified sensitive industry. The researchers used descriptive method with an independent model and chi-square test is used. They selected those companies which have same level of impact upon the environment. A test of no-response bias was undertaken. A Mann Whitney U test was used and null hypothesis was measured based on 0.05 level of significance. The researchers show some specific areas (costing system, budgeting system, capital budgeting and expenditure, investment appraisal, performance measurement and appraisal, internal reporting mechanisms, risk assessment etc.) within the management systems whether these are existing in a firm or not. The researchers found that only costing system creates difference in firm to firm. They observed that environmental costs were not being appropriately identified dividing the firms into sensitive and less sensitive. They identified some issues for analysis such as waste, energy usage, recycling, returnable packaging, pollution, accounting for rehabilitation, environmental contingent liabilities, life cycle cost analysis in product development, environmental costs in production costs and addressing legal regulations. The researchers found that there was a significant difference for the adoption of cost benefit analysis of site contamination (0.01) and site cleanup (0.05). They asked the respondents about general environmental audit, waste audit, energy audit. The respondents told that greater proportion of firm which are in environmentally sensitive industries conducted general environmental audit. The researchers discuss about environmental reporting. They find that sensitive industries adopt environmental reporting practices by firms to broader community.

The analysis found that for environmental issues, there was no significant difference in the level of adoption of environment-related management accounting practices. Environmental sensitive firms are more likely to develop environment-related management accounting procedures. Finally, the researchers suggested that further research is needed to identify what causes organizations to adopt environment-related management accounting procedures (Frost & Wilmshurst, 2000).

**Schaltegger et al., (2000)** in their paper "Environmental Management Accounting: Overview and Main Approaches" tried to combine EMA's two concepts such as internal environmental accounting using a monetary measurement and monetary and non-monetary measurement. At first, to define EMA, the authors explain briefly about these two approaches of EMA. In First approach, they told that environmental accounting is the combination of environmentally differentiated conventional accounting & ecological accounting framework of environmental accounting (see Schaltegger & Burritt, 2000, 58ff; Schaltegger et al. 1996, 12ff) and corporate bases of environmental awareness (Alam & Zakaria, 2013). The authors also define EMA as general term for corporate internal environmental accounting. It includes both monetary and non-monetary internal accounting approach. In the approach, three aspects are described that are management accounting (conventional management accounting, environmental management accounting), financial accounting (conventional financial accounting, environmental financial accounting), other accounting systems (other conventional accounting systems, other environmental accounting system).

Based on the environmental accounting system framework (Burritt, Hahn and Schaltegger, 2002) further developed the EMA framework which considers four information needs of making management decisions. Here, EMA combines two aspects monetary environmental management accounting (MEMA) and physical environmental management accounting (PEMA. Though different authors distinguish these two approaches of EMA, this article concludes that monetary and physical accounting systems are necessary to managers in seeking to reduce environmental impacts of organizations. The authors of this article also say that EMA is necessary for effective scientific communication and research as well as promotion and establishment of modern EMA approaches in practice (Schaltegger et al., 2000).

**Burritt (2005)** conducted a research on "Challenges for Environmental Management Accounting" to identify the potentiality of EMA and consider the meaning of the development of EMA. This article includes some reasons of being interested in EMA information i.e. control and reduce environmental costs by management; improve income or profitability or environmental impacts (to reduce penalties); promotion of EMA; EMA tools; These tools usage are full cost accounting or life cycle costing, environmental cost management, material flow cost accounting, environmental capital appraisal, environmental capital performance appraisal. Here, the paper discusses about EMA's development using six possible term to measure the usefulness, to internal and external stakeholders, and to provide monetary and physical unit management. The author identifies some key problems with conventional management accounting. Problems with conventional management accounting are narrow performance appraisal techniques and their short term focus, less attention to articulation of stocks and flows, less focus on manufacturing. Problems with lack of recognition of environmental impacts in conventional management accounting are less importance on environmental costs, some environmental costs items are not tracked and traced, indirect environmental costs are included with general business overheads, investment appraisal excludes environmental considerations, flow accounting for sustainability issues. Burritt also reveals challenges for EMA to address inductive theory and the directions of case studies, not focusing small and medium enterprises and enterprises in developing countries, not focusing win-win outcome, pure physical information is not considered as environmental management accounting information, software systems, distinction between internal and external stakeholders' usefulness, performance management and appraisal system, problems of cost allocation. The paper concludes that relevant, reliable, low cost EMA information is needed if the impetus already started is to continue to gather pace (Burritt, 2005).

**Cerin and Laestadius (2005)** in their study "Environmental Accounting Dimensions: Pros and Cons of Trajectory convergence and increased efficiency" focused on the internal incompatibility of the development of EMA systems dimension related to environmental accounting. Here, three dimensions of environmental accounting practices and accompanying management fields are indicated. These are regional, company and product accounting. The researchers identify regional dimension of environment management focusing on intra and inter-regional materials meaning EMA dimension exchange across the border. The researchers discourse Judicial (municipal/country) accounting, Hinterland Accounting, Metal and Accounting. Secondly they identify a company dimension in environmental management and accounting. This includes two major discourse judicial (company) accounting, Value Chain Accounting. Thirdly, the researchers identifies product dimension in environmental management/ accounting. They showed some barriers to intra and inter-environmental accounting such as the adequate data is limited continuously, accounting dimensions are incompatible, delimitation of the various accounting dimensions are overlapped. To remove such barriers, the authors have not been suggested new alternative. They have said that existing systems need to be developed further. They also have said that using I/O analyses of the three dimensions of environmental accounting helps to cut cost for the most resource consuming parts of this study. They have argued that the usage of various assessment paths may be useful and contribute to build compatibility between these dimensions of environmental accounting to estimate environmental impact (Cerin and Laestadius, 2005).

**Pohjola (2005),** conducted a study "Applications of an Environmental Modeling system in the Graphics Industry and Road Haulage Services" to explore the Application of an Environmental Modelling System in the Graphics Industry and Road Haulage Services. He designs this article to identify, analyse, manage and reports environmental factors related to operational and financial functions in business processes. He says that environmental management is an increasingly important part of strategic management in companies. The environmental pollution caused by industrial countries especially air emissions. He develops a framework based on the method of environmental accounting (Pohjola, 2005).

**Kokubu & Nashioka (2005)** conducted a study on "Environmental Management Accounting Practices in Japan" to show the scenario of environmental management accounting and different initiatives in industry operations. They argued that environmental accounting practice in Japan had been led by two governmental initiatives. One is MOE (Ministry of the Environment) initiatives. Another was METI (Ministry of Economy, Trade, and Industry) initiatives. MOI initiatives focused on external disclosure (such as provide information for external stakeholders and public). METI initiatives focused on internal applications of environmental accounting which is EMA (Kokubu & Nashioka, 2005)

**Venturelli and Pilisi (2005)** in the project paper "Environmental Management Accounting in Small and Medium-sized Enterprises: How to Adopt Existing Accounting Systems to EMA Requirements" focused on how existing accounting systems of the companies involved have been modified to allow a regular monitoring of environmental costs in Italian and Brescian small and medium-sized enterprises. There are 60 SMEs selected as sample size according to ISO 14001 standards. The writers find that 90 percent of industrial companies have been 10 and 50 employees and 95 percent have less than 100 in both Italy and Brescia. They develop a joint checklist and provide it to the companies to collect environmental costs and quality costs. They show different environmental costs in their project paper (prevention cost, monitoring costs and failure costs). They also show how environmental costs are distributed. They find that failure costs have been liable and passive attitudes towards environmental issues: consequently companies are not naturally attracted by EMA. The writers select three cases to measure the successful implementation of EMA. In case 1, company with 100 employees took only environmental costs as directs cost (i.e. tax and packaging). But company added an environmental costs centre to collect environmental costs using seven new accounts to the company's accounting system to address environmental costs (purchase of materials for EMS, consulting for EMS, External audit for environment and quality, purchase of materials for waste water plant, waste analysis, environmental maintenance or operational costs, waste disposal. Here the waste disposal is only considered as directs cost. In case 2, the company selected five new accounts. Those were consulting for EMS, purchase of materials, waste disposal and transport, environmental training for personnel, environmental maintenance, waste taxes, and environmental penalties. In case 3, company with 150 employees collected environmental costs in the existing quality costing system. These cost are quantified in the year of ISO 14001 certification. The company used to calculate the products cost plus cost of raw materials to direct personnel cost to other costs. This article has presented three cases of SMEs which had no system for measuring and analysing environmental costs. Following the guidelines of AIB, these companies have set up EMA to quantify their environmental costs and set a suitable accounting system to record those (Venturelli & Pilisi, 2005).

**Hyrslova and Hajek (2005)** in their study "Environmental Management Accounting in the Framework of EMAS II in the Czech Republic" deals with requirement on EMA implantation in the framework of EMAS as well as reporting environmental costs and revenues. It shows that the enterprises in Czech Republic decrease the detrimental impact on the environment by introducing environmentally-friendly technologies. The study is qualitative and quantitative in nature. In 2002, a survey was conducted with a questionnaire to carry out on analysis concerned with the purpose for introducing EMS in Czech enterprises where 208 enterprises introduce EMS at the time of the survey, from them 89 companies participated in the survey. This survey found that companies were interested to introduce EMS for protecting the environment and attempts to improve the company's position in the market. Most companies include environmental targets in their company strategies. If companies want to implement EMAS, then they need to develop a system, to calculate environmental costs in order to implement environmental management accounting (EMA). If EMA system information use properly, a reduction in the environmental impacts of the activities, products and services of the enterprises, a reduction in environmental risks and an improvement in the economic results of the enterprise will be achieved. The study also confirm that implementation of EMA does not constitute a major problem for the enterprise. Some employees' attitudes indicate an unwillingness to introduce any changes in existing functioning information systems (Hyrslova and Hajek, 2005).

**Lee et al. (2005)**, in their study "Environmental Accounting Guidelines and Corporate cases in Korea: Implications for Developing Countries" focused on introducing the Korean environmental accounting guideline (draft) corporate cases. They discussed about main issue to adopt environmental accounting successfully into companies in Korea as well as in developing countries. Most companies in developing countries are still far behind in introducing and using environmental accounting techniques and methods compared to advanced companies in developed countries. The paper also find that the pressure of external stakeholders like financial institutions, social responsible investment, the Government and local communities are responsible to lead the companies to show interest in environmental accounting. The introducing guidelines suggest the activity based environmental costing approach to the company to classify environmental cost. The guideline define environmental costs as resources and develop a disclosure format of those costs – pollution treatment activity costs, pollution prevention activity costs, stakeholder relation activity costs, environmental compliance and remediation activity costs. The paper describes two company cases in Korea (Lee et al., 2005).

**Epstein & P.S. (2001)** in their paper on "Using a Balanced Scorecard to Implement Sustainability" offered advantages for separating environmental and social performance indicators from the other perspectives and into an added fifth perspective. Having a stand-alone perspective reflects the importance of environmental issues to a company's strategic objectives, and forces managers to focus more on the metrics separately than if they were embedded with other metrics in different perspectives. The notion of managers giving more weight to environmental metrics in a separate perspective can be predicted based upon principles from Gestalt psychology (Epstein & P.S. 2001).

**Marcus (2005)** in his study "Environmental Performance and the Quality of Corporate Environmental Reports: The role of Environmental Management Accounting" attempted to analyse whether there is an association between environmental performance and corporate environmental reporting in the paper and electricity industries in Germany and the United Kingdom. The main objective of this study is to assess what extent the level of physical environmental performance and the quality of its corporate environmental reports in two industrial sectors and two EU countries. At first, the study discusses about environmental performance and measurement system and also discusses about environmental reporting. Environmental Performance Measurement (EPM) is the measurement of the interaction between business and the environment (Bennett & James, 1997). Corporate Environmental Reports (CERs) are "stand-alone reports issued by companies to disclose Environmental information available to the public" (Brophy & Starkey, 1996). CERs have different potential users such as businesses, financial institutions, consumers, communities and government agencies; this makes different reporting requirement necessary (Bennett & James, 1998). Then, a measurement framework is introduced for identifying the empirical findings. To operationally measure the environmental performance (according to ISO 14031 as defined in ISO, 1996, 1999) and the quality of environmental reporting (based on the criteria mainly developed and proposed by IRRC, 1995), three groups of variables have been selected. These are general variable, Environmental Performance, Environmental Reporting. The survey is conducted among the firms in the electricity and paper industries in the United Kingdom and Germany to analyze possible association between environmental performance and environmental reporting. The two sectors and countries are selected because of their environmentally intensive sectors as well as their location. The study uses exploratory data analysis of air emissions and quality of environmental reports. These findings suggest that consistency between environmental performance and environmental reporting using the terms of statistical correlation. Finally, the study finds out that Environmental Management Accounting tools may have an important influence on linking the quality of environmental reporting to the environmental performance of firms (Marcus, 2005),

**Scavone (2005)** conducted the study "Environmental Management Accounting: Current Practice and Future Trends in Argentina" to identify environmental management accounting techniques in projects carried out by Argentina, South America. The focus was on reflecting environmental factors in order to make a sustainable contribution to both business success and sustainable development. He stated that cleaner technology along with EMA adds value for measuring environmental impacts and cleaner production strategy will drive for new product concepts, new productions processes, and logistics solutions. Some key considerations were given to the benefits and barriers of co-operation between local government and the private sector. The paper conclude that for any environmental or social phenomena, there are many variable have a direct cause-effect relationship and link between achievement of programs and the advertised results. The paper states that cleaner technology along with EMA adds value for measuring environmental impacts and cleaner production strategy will drive for new product concepts, new productions processes, and logistics solutions. Some key considerations are given to the benefits and barriers of co-operation between local government and the private sector (Scavone, 2005).

**Yeshmin (2015)** in her study on "A Study on Cost and Management Accounting Mechanism as Practiced in Manufacturing Industry of Bangladesh" expected to explore the practice of cost and management accounting mechanism of manufacturing industry in Bangladesh. She surveyed 66 manufacturing organizations to complete sample size. Structured questionnaire was used to conduct the study using 5 point Likert Scale measurement. She developed a checklist of 49 cost and management accounting mechanisms. She found that budgetary control is the most used mechanism used by the sample out of 49 mechanisms. She also found that Kaizan costing, six sigma and value engineering etc. were not familiar to the sample as new mechanism. Findings reveal that out of forty nine cost and management accounting mechanisms, budgetary control is the most frequently used mechanism by the sample. The study has also recognized 12 components that had explained 79.862 percent of the total variation in the application level of cost and management accounting mechanisms. She thought that these findings will help to develop the perception about the cost and management accounting (Yeshmin, 2015).

**Montabon et al. (2007)** in their study "An examination of corporate reporting, environmental management practices and firm performance" aimed to explore the scenario of the firms whether they practice environmental management, environmental reporting or not and their performance. The authors used an innovative data Sources to explore environmental management practices in order to test the relationship between environmental management practices and firm performance. The secondary data were collected from 45 corporate reports related to environmental and business performance data. Content analysis was used to gather data and canonical correlation was used for analysis in a two-step process. They added some issues of environmental operational practices such as recycling, waste reduction, remanufacturing, consume internally, packaging:, spreading risks, creating a market for waste products, energy, environmental information, rewards as incentives as environmental project. They also added some variables of environmental practices such as Tactical, Strategic, and Performance measures in order gather primary data. They found that there was a significant positive relationship between EMPs and firm performance. They stated that there was a lack of standards practices in environmental practices by firms (Montabon et al., 2007).

**Khanna & Anton (2002)** made a study on "Corporate Environmental Management: Regulatory and Market-Based Incentives" aimed to identify and measure the status of corporate environmental management. Survey method was used and 500 firms were selected as sample to collect data. They found that threat of environmental liabilities, high cost of compliance, market pressures, and public pressures on forms with high on-site toxic emissions per unit output create incentives for adopting a more comprehensive environmental management system (Khanna & Anton, 2002).

**Masanet-Lodra (2006)** in his paper "Environmental Management Accounting: A Case Study Research on Innovative Strategy" aimed to conduct a study on environmental management systems developed in the ceramic tiles sector. The aim of his study was to show the relationship between firms and environment, analysing environmental positions in companies assumed in their environmental strategy and their environmental behavior reflected in facts. Case study research methodology was applied to this study. The researcher found that elaboration of a larger amount of environmental accounting information for internal use than external use. The researcher also confesses the limitations of case study methodology that means it creates findings to theory thesis but not for whole population basis (Masanet-Lodra, 2006).

**Islam & Dellaportas (2011)** made a study "Perceptions of corporate social and environmental accounting and reporting practices from accountants in Bangladesh" to know the accountants' perceptions regarding corporate social and environmental accounting and reporting practices in a developing country like Bangladesh. They told that Survey and Interview method was used to collect data from the members of the Institute of Chartered Accountants of Bangladesh. They found that accountants had a positive perception towards corporate social and environmental accounting and reporting practices (Islam & Dellaportas, 2011)

**Gauthier (2005),** in his study "Measuring Corporate Social and Environmental Performance: The Extended Life-Cycle Assessment" aimed to combine business ethics to corporate social responsibility with social and environmental dimensions. The authors wanted to identify corporate environmental management tool such as Life Cycle Assessment in order to clear integration of social performance in corporate activities. Qualitative and quantitative methods were applied in the study. The author selected product life-cycle method firstly. He added various stages of product life-cycle methods such as extraction of its raw materials, manufacture, packaging, storage, distribution, use, and recycling-destruction. He state some environmental criteria used in each stages registered in a spreadsheet were consumption of energy, consumption of raw materials, consumption of  water, production of polluting agents, production of toxic products, production of waste. The author suggested that widely used LCA tool be extended for the purpose of sustainable development. Because, LCA assists the company to edit or add its social or environmental yearly reports (Gauthier, 2005).

**Ramesh and Madegowda (2013)** made a study on "Environmental Accounting Practices in selected Indian Companies" aims to measure the performance of environmental accounting in Indian manufacturing companies towards environmental safety and welfare of people. They mentioned that Indian companies face two problems and trade-off between environmental protection and development is required. They stated some factors of the development of environmental accounting. These are improvement of environment for sustainable development; well established accounting guidelines are required. So they realized to evaluate the procedure followed by the selected companies as how these companies environmental costs and benefits and report the same to the stakeholders (Ramesh and Madegowda, 2013).

**Montabon et al. (2006),** made a study on "An examination of Corporate Reporting, Environmental Management Practices and Firm Performance" in order to develop and present a more comprehensive set of environmental accounting practices. The authors also aimed to examine the relationship between environmental Management Accounting and firm performance. They used the data in the paper collected from 45 corporate reports. They involve 7 institutes and 13 investors in data collection. Content analysis was used to gather the data. For the two step process analysis, canonical correlation was used. They conduct factor analysis which was related to canonical correlation to create the composite of variable. Conducting this cross-sectional analysis, they identified 4 dependent variables (product innovation, process innovation, ROI, sales growth) and 20 independent variables into 3 categories i. e. operational practices, tactical practices and strategic practice. Recycling, Proactive waste reduction, reactive waste reduction, remanufacturing, consume internally, market for waste, Money spent on environment were represented operational practices. Tactical strategies variable were early supplier involvement, environmental standards for suppliers, environmental audits, life cycle analysis, environmental design, specific design target, environmental risk analysis. Strategic practices were corporate policy, environmental mission statement, environmental department, surveillance of market, and strategic alliance. They measured the significant environmental management practices through these variables. They conclude that the study used a wide range of environmental management practices and these are positively associated with multiple of firm performance measures (Montabon et al., 2006).

**Norsyahida, Jusof, & Zulkifli (2016)**, in their article named "The Characteristics of the Company and Implement Environmental Management Accounting" include five features of each company. These were- sensibility to environmental issues industry, company size ownership status, acceptance environmental management and the percent of outside directors and its impact on the implementation of EMA. They emphasized more on environmental activities benefit and cost considerations. They found that extent of implementation of environmental accounting between different companies with different characteristics not much different with the exception of property ownership status (Norsyahida, Jusof, & Zulkifli (2016).

**Research Gap**

Thorough reviews of the previous researchers done in the area related environmental management accounting [Klassen and Mclaughlin (1996), Mclaughlin (1996), Frost & Wilmshurst (2000), Schaltegger et al. (2000), Epstein & P.S. (2001), Khanna & Anton (2002), Burritt (2005), Cerin & Laestadius (2005), Gauthier (2005), Hyrslova and Hajek (2005), Kokubu & Nashioka (2005), Lee et al. (2005), Marcus (2005), Pohjola (2005), Scavone (2005), Venturelli & Pilisi (2005), Masanet-Lodra (2006), Montabon et al. (2006), Sendroiu et al. (2006), Montabon et al. (2007), Islam & Dellaportas (2011), Ahmed (2012), UWUIGBE (2012), Makori & Jagongo (2013), Ramesh and Madegowda (2013), Tsui (2014), Jamil et al. (2015), Norsyahida, Jusof, & Zulkifli (2016), Yeshmin (2015), Karimi et al. (2017), Noah (2017)] are made these researchers. Almost all the studies focus on Social Performance, Environmental Performance, Corporate Environmental reporting, Environmental Performance Measurement, Physical Environmental Performance, Environmental Awards, Environmental Modelling System, Corporate Environmental Disclosure, Sustainability Development etc. Very few studies have been conducted on Environmental Reporting in the context of Bangladesh (Dutta & Bose, 2008). But there is still a lack of scarcity of theory driven research on overall EMA practices on cement industry in Bangladesh, though Bangladesh is one of the most vulnerable countries because of environmental changes throughout the world.

Moreover, there is a lack of genuine knowledge about the factors related to the practices of EMA in manufacturing sector of Bangladesh. No comprehensive study on EMA practices has been made so far with reference to corporate sector in Bangladesh. Considering this gap, the study has been conducted to develop a conceptual framework after selecting variables of 13 categories based on GRI Guideline and previous literature to find out corporate Environmental Management Accounting Practices in Bangladesh in the next chapter.

**CHAPTER THREE**

**CONCEPTUAL FRAMEWORK OF THE STUDY**

**Introduction:** In the previous chapter we have presented the introductory aspects of the study and review of literature. In this chapter an attempt is taken to present the theoretical framework of the study**.**

**3.1.Definition of Environmental Management Accounting**: Environmental Management Accounting is defined using six possible term such as environmental, management, accounting, environmental management, management accounting, environmental management accounting. EMA is derived as a field of Social Accounting combining environmental Accounting and Environmental Management. These terms are discussed to measure the usefulness, to internal and external stakeholders, and to provide monetary and physical unit management for fulfilling the need for environmental accounting (Burritt, 2005). Here, some aspects of EMA discuss below:

**3.1.1 Social Accounting:** The term 'Social Accounting' is used as the name Social performance information, social audit, social accounting, socio-economic accounting, social responsibility accounting and social and environmental reporting have been used interchangeably in the literature. The research area of the study is related to environmental management accounting practices and it is the extent of social accounting.

**3.1.2 Environmental Accounting:** Environmental Accounting is a term that used broadly in the context of different meanings and applications (Shil & Iqbal, 2005). There were so many research works conducted on environmental accounting. The first environmental accounts were introduced by Norway in the 1970s and then these concepts were slowly adopted by other countries (Shil & Iqbal, 2005). After the environmental accounts introducing year, many developed countries, such as UK, USA, Canada, Japan etc. conducted research on environmental accounting information and these countries have taken many important decisions for their countries' firms about environmental protection (Zhang et al, 2009).

Alewine & Stone (2010) stated that Environmental Accounting information is being unique and modern metrics for facing environmental challenges (e.g. unfamiliarity) when these data are combined with traditional financial metrics in decision making. Because, it possesses the ability to provide accurate information in the financial statements about estimated social cost that are occasioned by the production externalities on the environment. It also measure how much deliberate intervention cost had been incurred to bridge the gap between the marginal social cost and the marginal private costs by a firm (Makori & Jagongo, 2013). Environmental Accounting instruments which are integrated into a company's information can lead to advantages such as increased quality of information and higher transparency within the enterprise. Thus, it can be stated that environmental accounting is the combination of environmentally differentiated conventional accounting & ecological accounting framework of environmental accounting (Schaltegger et al., 2000, Schaltegger & Burritt, 2000, 58ff; Schaltegger et al. 1996, 12ff).

**3.1.3. Conventional accounting:** It covers three aspects of accounting which are- conventional monetary accounting, conventional financial accounting, and other conventional accounting.
 Conventional monetary accounting is the central tools and is not disclosed to external stakeholders of sensitivity and confidentiality of information. Some issues i.e. Environmental cost and how they track and trace, how they treated, treated as overhead costs, environmental responsibilities of management accountant are discussed in it . Conventional financial accounting provides information which is related to the financial impact of firms' corporate activities to external stakeholders. Here some issues of financial accounting are discussed i.e. environmentally-driven outlays are recorded as assets or expense, environmental liabilities related information is disclosed using standards and guidelines or not, recommendations for treating techniques to handle these liabilities, environmental aspects and their measurement, emission trading certificate is included in financial statement or not. Other conventional accounting combines two aspects such as tax accounting and bank regulatory accounting (Schaltegger et al., 2000).

**3.1.4. Ecological accounting:** Ecological accounting can also be classified into three systems such as (i) internal ecological accounting, (ii) external ecological accounting, (iii) other ecological accounting. Internal ecological accounting aims to collect information about ecological system for management's internal use and this information are expressed as physical units. External ecological accounting aims to collect and provide data on environmental issues to external stakeholders such as general public, communication media, shareholders, environmental funds, non-governments organizations and pressure groups. Other ecological accounting aims to provide a means for regulations to control compliance with regulations (Schaltegger et al., 2000).

**3.1.5. Environmental Accounting Practices Issue:** Some EMA Practices issues are-
(i) Accounting for wastes; (ii) Accounting for energy usage; (iii) Accounting for recycling
(iv)Accounting for returnable packaging/containers; (v) accounting for pollution; (vi)
Accounting for land remediation; (vii) Accounting for environmental contingent liabilities; (viii)
Accounting for costs of legal regulations (ix) Accounting for sustainability (Burritt et al., 2000).

**3.1.6. Three Dimensions of Environmental Accounting and Management:** Bennett et al. (2002b, p.2) emphasized that EMA provides a close link between environmental management and management accounting through the introduction of a new system that reflects a change in management philosophy towards concern for the environment as an ongoing issue for business. The emphasis is mainly given upon as the basis for considering corporate environmental issues (Howes, 2002; Bennett et al., 2002b, p.2). Cerin & Laestadius (2005) is identified three dimensions such as- (i) Regional Dimension of Environmental Management; (ii) Company Dimension of Environmental Management;(iii) Product Dimension in Environmental Accounting/Management

**3.1.7. Environmental Management Accounting:** Environmental management accounting (hereafter EMA) grew up from corporate environmental accounting and branched off along the lines of management accounting (Afzal, 2012). It has emerged during the last two decades in response to fulfill the requirement of environment friendly initiatives and issues adopted by the interested parties for sustainable development. This is because of it has a strong focus on providing better information on the actual environmental costs already incurred by the entity to the management of an organization (Shil & Iqbal, 2005). It provides a broad set of principles and approaches that provide the materials/energy flow and cost data needed many other environmental management activities and programs. While EMA data can be used for most types of routine decision-making within an organization, the data are particularly valuable for management initiatives with a specific environmental focus (Lundholm & Myers, 2002). That why, it is recognized that EMA has developed because of the limitations of management accounting (Tsui, 2014). EMA includes both monetary and non-monetary internal accounting approach.

**3.1.8. Information Types of EMA:** There are two types of information are found from EMA implementation. These are physical environmental/non-monetary information and monetary information discussed in below:

**(i) Non-monetary information related to EMA:** The information of material, flow of energy, water, waste management, emissions etc. are non-monetary environment-related information. More specifically the information of the amount of material used, volume of fresh water consumed, the volume of wastes generated, energy consumption, air emission etc. and their business environmental impact in physical units measurement (Burritt et al, 2002; Shen & Tam, 2002; IFAC, 2005; Venturuelli & Pilisi, 2005; Afzal, 2012; Uwuigbe, 2012; GRI, 2013; Norsyahida, 2015).

**(ii) Monetary information related to EMA**: Monetary environment-related information makes a relation to environmental costs and earnings. Elaborately, effluent and waste control costs, emission control costs, sales from scrap and wastes, recycling subsidies and tax incentive on green equipment, material conservation costs, remediation costs, development cost of energy efficiency, control-regulatory compliance costs etc. are included in monetary environment related information. (Burritt et al, 2002; Shen & Tam, 2002; IFAC, 2005; Venturuelli & Pilisi, 2005; Afzal, 2012; Uwuigbe, 2012; GRI, 2013; Norsyahida, 2015). Monetary environmental information can also be referred to financial figure or representation of physical environmental information.

**3.1.9. Technological Aspects of EMA***: EMA adoption with managerial technology is related putting knowledge to practical use (Burgelman and Muidique, 1988). The term Managerial Technology as a field of EMA is derived as the sector of Managerial Technology in which EMA combines Knowledge, methodology and practice of Technology and applies these to combine environmental management and economic results. There are some tools and techniques of information collection, targeted information analysis and communication are covered by EMA regarding Managerial Technology. These tools and techniques are used to mediate between inputs and outputs (Tushman and Anderson (1986, p.440) as cited by Abrahamson, 1991). Some companies decide to adopt 'Managerial Technology' in the field of EMA if it is relevant to them.

**3.1.10. Environmental Costs:** Lack of standard of Environmental costs in implanting EMA causes the major problem in this area. There are various internal costs as well as external costs incurred inside and outside of the company such as disposal costs or investment costs which causes damage and hinders environment protection. Cost of wastes, land rehabilitation costs and R&D expenditure on green initiative these are directly related to the product and/or services (Jasch, 2003). Companies are directly liable for these costs which cannot be avoidable (De Beer and Friend, 2006). On the other hand, externalities refers the external costs that companies are not legally accountable for because they are financially immeasurable and attributable (Jasch, 2003; 2006; Crowther and Aras, 2008; Jones, 2010).

Thus, it hampers to the mass people. The effect of environmental costs in cement companies are also a tremendous problem. Environmental costs saving are crucial factor for profits of corporate environmental activities. Most of environmental costs are usually not traced in a systematic way but it can be summed up in general overheads**.** The aim of environment protection project is to prevent and omit disposal substances in order to utilizing raw materials and its other supporting objects without any hamper to the environment. To project the environmental costs it is need to add costs of wasted materials with the costs of wasted capital and labor by which total corporate environmental costs can be found (Aliu, 2011). There are some environmental costs indicators are found on previous research such as- purchase of materials for EMS, purchase of control instruments, monitoring of audit emissions to air waste water, noise measurement, waste analysis, operational costs of aspiration and abatement systems, communication for environment, waste disposal, training for environment, penalties/fines so on (Venturuelli & Pilisi, 2005).

**3.11. Barriers of EMA Practices:** All efforts of environment friendly production process, operation, management will be failing if proper promotion is not occurred. Inadequate promotion is a great barrier for EMA practices. Burritt & Saka (2006) discovered from the survey on Japanese companies that the practice of eco-efficiency measurement with EMA information was incomplete and eco-efficiency information was underutilized for lack of adequate public promotion. EMA practices & information must require collaboration and actions by different functions, such as financial management and environmental management. But there is no logical relation established between them (Bartolomeo et al., 1999).

**3.12. Representation of Environmental Information:** Environmental Information must be disclosed according to ISO standards and GRI guidelines. Generally a very little organization reported their environmental information previously. But things have been changed. There is need arises in reporting their business environmental impacts or environmental performance in their (Nik Ahmad and Sulaiman, 2004; Yusoff et al., 2007; Alrazi et al., 2009). Any kind of unawareness and unwillingness in environment-related information disclosure can be attributed to the creation of companies' environmental information to attain legitimacy (Sulaiman and Nik Ahmad, 2006). From earlier literature review, it is found that the accountants' involvement in disclosure is limited and they are not so very interested to disclose all the information related to environmental aspects (Lodhia, 2003; Collins et al., 2011). Prior studies found that most companies reported significantly more narrative and positive environmental information with lack of concentration on specific environmental information concerning the environmental impact of business activities, environmental awareness, and related monetary implications of such information (Hackston and Milne, 1996; Ahmad et al., 2003; Alam & Zakaria, 2013; Ferreira, 2004; Nik Ahmad and Sulaiman, 2004; Jaffar, 2006; Yusoff et al., 2007; Alrazi et al., 2009; Buniamin, 2010; Bouten et al., 2011).

The study investigated the influential factors in the organizational, environmental and technological contexts on firms' intentions to the practice of Environmental Management Accounting. Firms with higher profitability are motivated to disclose information to favorably distinguish themselves from other firms (Dye, 1985, Kaur and Lodhia, 2014; Ho and Taylor, 2013).

### 3.13. Environmental Management Accounting and Reporting Guidelines, Principles, and Standards

**3.13.1 GRI Principles related to Environmental Aspects:** Global Sustainability Standard Board takes GRI for the organizations to disclose information under some criteria. There are some specific standards disclosures on Management Approach indicators. These are Economic, Environmental, and Social category. Social category includes some sub-category like, Labor Practices and Decent Work, Human Rights, Society, Product Responsibility. The study mainly focuses on GRI principles which discuss about Environmental Management Accounting Practices Indicators broadly and these principles are suggested to practice by all manufacturing or other corporate organizations mandatorily for the betterment of environment. A group of twelve environment-related aspects are discussed in these guidelines. These are Material, Energy, Water, Biodiversity, Emissions, Effluent and Waste, Products & Services, Compliance, Transport, Overall Environmental Costs, Supplier Environmental Assessment, Environmental Grievance Mechanisms, and another category is Health and Safety (GRI, 2013). These are related to disclosure of environmental approach.

Various specific Standards under these twelve categories are also discussed here. In Material category, material used by weight or volume (Non-renewable and renewable material used) (G4-EN1), percentages of material used that are recycled input materials (G4-EN2). In Energy category, energy consumption within and outside the organizations, energy intensity, reduction of energy consumption, reduction in energy requirements of products and services are discussed (G4-EN3 to G4-EN7). In water, total water withdrawal by source, water sources size, percentage and total volume of water recycled and reused (G4-EN8 to G4-EN10). Biodiversity aspects are operation sites information whether these are owned, leased or managed to protected areas or not, geographical location of sites, operation types, size of operation at site, impact of products on Species, extent of areas affected duration of impacts of affection etc. Emission aspects included in the Guidelines are air emission, direct CHG emission created from organization's operations, Energy emission from electricity, heating, cooling, steam, other indirect emission etc. (G4-EN15 to G4-EN21). Effluent and waste category includes the variable like total water discharge, total weight of hazardous and non-hazardous waste reuse, recycle, burn of waste onsite storage, hazardous waste transported, percentages of hazardous waste shipped (G4-EN22 to G4-EN26). In products and services, impact mitigation of Environmental practices of products and services, percentage of products sold and their packaging material are discussed. Compliance category measures the number of compliance and monetary value of fines with environmental laws and regulations. In transport aspects, significant environmental impacts of transporting products, and other goods and materials for the organization's operation are included. GRI guidelines include allocation of overall environmental costs. These are environmental protection costs of waste disposal, costs of emission treatment, remediation costs, prevention and environmental management costs. Supplier Environmental Assessment involves information of percentage of new suppliers that were screened using environmental criteria, environmental impact on supply chain. These are some grievance mechanisms included in these guidelines showing environmental impacts on organizations. Health and Safety issues are also included in environmental aspects.

**3.13..2 Links with United Nations Global Compact Principles 2000 about Environmental Aspects:** According to the principles of United Nations Compact (2000), there are some attributes of environmental practices in worldwide for measuring their condition about disclosing environmental information to stakeholders who are related to the enterprise internally and externally. In principle 1, it is included that business should support and respect the protection of internationally proclaimed human rights. This principle is related to human rights, society, and local community. In principle 2, business should not involve in those activities which are harmful for human and violate human rights. Principle 3 is related to Labor practices and decent work i.e. labor/management relations, and freedom of Association and Collective Bargaining (G11). Principle 4 is related to upholding the elimination of all forms of forced and compulsory labor and principle 5 describes upholding the effective abolition of child labor. Principle 6 is related to non-discrimination in respect of employment and occupation. There are three principles closely related to environmental aspects stated below: Principle 7: Business should support a precautionary approach to environmental challenges. Principle 8: Business should undertake initiatives to promote greater environmental responsibilities. Principle 9: Business should encourage the development and diffusion of environmentally friendly technologies. (UN Global Compact Principles, 2000).

**3.13.3 Links with OECD Guidelines for Multinational Enterprises about Environmental Reporting, 2011:** There are some beneficial determinants of Environmental management Accounting Practices in Multinational Enterprises are introduced G4 sustainability reporting guidelines and global reporting initiatives issued by Global Sustainability Standards Board. These are included in labor practices and decent work such as operational health and safety, training & education. These guidelines also include environmental aspects for benefit of local communities, impact measurement of environmental aspects on society i.e. supplier assessment impacts, Grievance Mechanism impacts, customer health and safety (OECD, 2011).

**3.14. Theoretical Framework of EA and EMA:** After the above discussion the following framework is developed by researchers for the study:

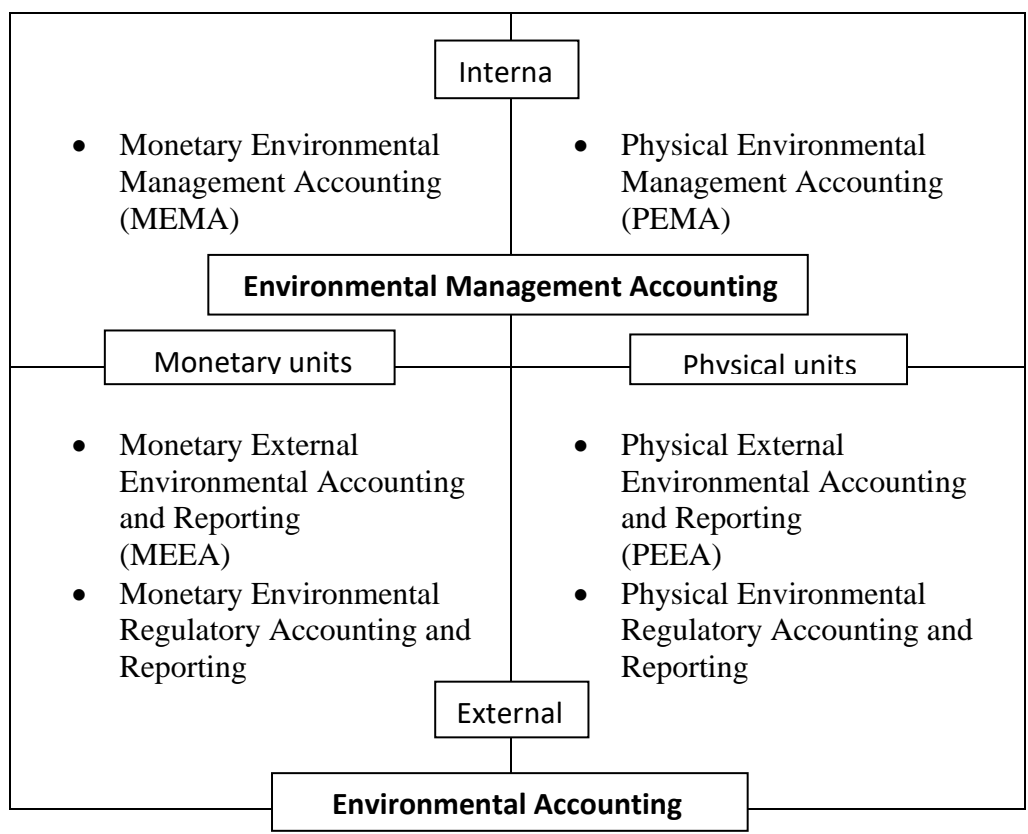

Figure 3.1: Categories of Environmental Accounting

(Modified from Bartolomeo et. al, 2000 p. 33)

**3.15. EMA Index:** The study develop an index of variable categorizing all into 13 group these are Material, Energy, Water, Biodiversity, Emission, Effluent and Waste, Products and Services, Compliance, Overall Environmental Costs, Supplier Environmental Assessment, Environmental Grievance Mechanisms, and Health and Safety. These variables are selected according to Global Reporting Initiatives Index and previous literature. The constructs used in this study are adopted and developed shown in below:

Table 3.1: Constructs and Item Sources

| Category | | Sources |
|---|---|---|
| **A. Material** | | Frost & Wilmshurst, 2000; Schaltegger et al., 2000; Jasch, 2003; Clarkson et al., 2008; Alrazi et al., 2009; Beck et al., 2010; Aliu, 2011; Afzal, 2012; Uwuigbe, 2012; GRI, 2013; Ramesh and Madegowda, 2013; Doorasamy, 2015; Norsyahida, 2015 |
| AM1 | Total weight or volume of material used | |
| AM2 | Non-renewable material used | |
| AM3 | Renewable material used | |
| AM4 | Packing material reserve | |
| AM5 | Conservation of material resources | |
| AM6 | Operating wasted material reserve | |
| **B. Energy** | | Frost & Wilmshurst, 2000; Schaltegger et al., 2000; Clarkson et al., 2008; Alraziet al., 2009; Beck et al., 2010; Aliu, 2011; Afzal, 2012; Uwuigbe, 2012; GRI, 2013; Ramesh and Madegowda, 2013; Doorasamy, 2015; Norsyahida, 2015 |
| BE7 | Energy consumption within the organization | |
| BE8 | Fuel consumption from non-renewable Sources | |
| BE9 | Fuel consumption from renewable Sources | |
| BE10 | Electricity consumption | |
| BE11 | Heating consumption | |
| BE12 | Cooling consumption | |
| BE13 | Steam consumption | |
| BE14 | Energy consumption outside the organization | |
| **C. Water** | | Frost & Wilmshurst, 2000; Schaltegger et al., 2000; Alrazi et al., 2009; Burritt et al., 2002; IFAC, 2005; GRI, 2013; |
| CW15 | Total water withdrawal | |
| CW16 | Water Sources size | |
| CW17 | Volume of fresh water consumed | |

| | | Norsyahida, 2015 |
|---|---|---|
| **D. Biodiversity Aspect** | | Clarkson et al., 2008; Alrazi et al., 2009; Beck et al., 2010; GRI, 2013; Norsyahida, 2015 |
| DB18 | Operation site owned, leased, managed to protected areas | |
| DB19 | Geographical location | |
| DB20 | Operation type | |
| DB21 | Size of operational site in km. | |
| DB22 | Impact of operation, products or service on species | |
| DB23 | Extent of areas impacted by operation | |
| DB24 | Duration of impacts | |
| DB25 | Rehabilitation of local community | |
| **E. Emission** | | Frost & Wilmshurst, 2000; Schaltegger et al., 2000; Clarkson et al., 2008; Alrazi et al., 2009; Beck et al., 2010; Aliu, 2011; Afzal, 2012; GRI, 2013; Ramesh and Madegowda, 2013; Norsyahida, 2015 |
| EE26 | Air emissions | |
| EE27 | Direct CHG omissions created by organization from operation | |
| EE28 | Energy omissions from generation of electricity, heating, cooling, stem. | |
| EE29 | Handling treatment and disposal of waste and emission | |
| **F. Effluent and waste** | | Frost & Wilmshurst, 2000; Schaltegger et al., 2000; Jasch, 2003; Clarkson et al., 2008; Alrazi et al., 2009; Beck et al., 2010; Aliu, 2011; Afzal, 2012; Uwuigbe, 2012, GRI, 2013; Ramesh and Madegowda, 2013; Norsyahida, 2015 |
| FE30 | Total water discharge by quality and destination | |
| FE31 | Total weight of waste by type and disposal method | |
| FE32 | Total weight of hazardous and non-hazardous waste | |
| FE33 | Reusing of waste | |
| FE34 | Incineration or mass burs of waste | |
| FE35 | On site storage. | |
| FE36 | Hazardous waste transported | |
| FE37 | Sales amount of scrap or waste | |
| FE38 | Recycling container and packaging | |
| **G. Products and Services** | | Clarkson et al., 2008; Alrazi et al., 2009; Beck et al., 2010**;** Afzal, 2012; GRI, 2013; Ramesh and Madegowda, 2013; Norsyahida, 2015 |
| GP39 | Environmental impacts consideration of products and services | |
| GP40 | Standard process of production | |
| GP41 | Quality of products and services | |
| GP42 | Packaging Accuracy | |
| **H. Compliance** | | Clarkson et al., 2008; Alrazi et al., 2009; Beck et al., 2010; Afzal, 2012; GRI, 2013; Norsyahida, 2015 |
| HC43 | Code of Conduct related to environmental aspects | |
| HC44 | Addressing legal rules and regulations | |
| HC45 | Training and education for employees, stakeholders about laws | |
| HC46 | Fines charged about environmental laws and regulations | |
| **I. Transport** | | Clarkson et al., 2008; Alrazi et al., 2009; Beck et al., 2010**;** Afzal, 2012; GRI, 2013; Norsyahida, 2015 |
| IT47 | Significant environmental impacts of transporting products and other goods and materials for the organization's operations. | |
| IT48 | Environmental impacts of transporting members of the organization's workforce are mitigated. | |
| IT49 | Cement supply and ship economically | |
| **J. Overall Environmental Costs** | | UN DSD, (1998); Frost & Wilmshurst, 2000; Schaltegger et al., 2000**;** Clarkson et al., |
| JO50 | Raw Material Conservation costs | |
| JO51 | Raw material purchase value | |

| JO52 | Packaging material costs | 2008; Alrazi et al., 2009; Beck |
|------|--------------------------|--------------------------------|
| JO53 | Toll manufacturing costs | et al., 2010; GRI, 2013; Ramesh |
| JO54 | Expenditure in Energy | and Madegowda, 2013; |
| JO55 | Fuel costs | Doorasamy, 2015;Norsyahida, |
| JO56 | Environmental Protection Costs of Waste Disposal | 2015 |
| JO57 | Site Restoration Costs | |
| JO58 | Water Consumption Costs | |
| JO59 | Depreciation costs for equipment | |
| JO60 | Costs of taxes, insurance permitted | |
| JO61 | Remediation and compensation costs for environmental damage | |
| JO62 | Development costs for energy efficiency products | |
| JO63 | Total monetary value of fines about environmental laws | |
| JO64 | Fines and penalties for incompliance of laws | |
| JO65 | Site Clean-up Costs | |
| JO66 | Transportation Costs | |
| **K. Supplier environmental assessment** | | UN Global Compact, 2000; |
| KS67 | New suppliers that were screened using environmental criteria. | Montabonet al., 2007; Clarkson et al., 2008; Alrazi et al., 2009; |
| KS68 | Environmental impact on Supply chain | Beck et al., 2010; GRI, 2013 |
| KS69 | Environmental Awards for suppliers | |
| **L. Environmental Grievance Mechanisms** | | GRI, 2013; Norsyahida, 2015 |
| LE70 | Grievances about environmental impacts filed, and addressed through formal grievance mechanisms. | |
| LE71 | Resolving the identified grievances during the accounting period | |

| M. Health and Safety | | OECD, 2011; Norsyahida, 2015 |
|---|---|---|
| MH72 | Safety measures applied for environmental protection | |
| MH73 | Cleaner Technology used | |
| MH74 | Occupational Health & Safety for employees | |
| MH75 | Training about safety | |

Source: Constructed based on GRI Guideline and Literature Review

# CHAPTER FOUR

## DATA ANALYSIS AND PRESENTATION

**Introduction**: In the previous chapters we have offered the introductory aspects, literature review and conceptual framework of the study. In this chapter we presented the analyses of data using the methods adopted in the study. In analyzing the data we presented firstly by industry to industry and then combined.

## 4.1. Analysis of data by Industry:

**4.1.1. Textile Industry:** There are 48 listed textile industries in DSE out of which 20 has been taken as sample for the study. The descriptive statistics of CEMRI and independent variables are given in table

Table 4.1 : The descriptive statistics of CEMRI and independent variables of sampled Textile Industries

| Descriptive Statistics | | | | | | | | |
|---|---|---|---|---|---|---|---|---|
| | N | Range | Minimum | Maximum | Mean | | Std. Deviation | Variance |
| | Statistic | Statistic | Statistic | Statistic | Statistic | Std. Error | Statistic | Statistic |
| TA | 5 | 17000.00 | 23000.00 | 40000.00 | 29800.0000 | 3056.14136 | 6833.73983 | 46700000.000 |
| TS | 5 | 11300.00 | 48700.00 | 60000.00 | 54600.0000 | 1827.29308 | 4085.95154 | 16695000.000 |
| BS | 5 | 3.00 | 12.00 | 15.00 | 14.0000 | .54772 | 1.22474 | 1.500 |
| SP | 5 | 21.00 | 39.00 | 60.00 | 48.6000 | 3.44384 | 7.70065 | 59.300 |
| CEMRI | 5 | 16.00 | 60.00 | 76.00 | 67.4000 | 3.62767 | 8.11172 | 65.800 |
| Valid N (listwise) | 5 | | | | | | | |

**Model Summary[b]**

| Model | R | R Square | Adjusted R Square | Std. Error of the Estimate | Durbin-Watson |
|---|---|---|---|---|---|
| 1 | .966[a] | .933 | .732 | 4.20169 | 2.123 |

a. Predictors: (Constant), SP, TS, BS;

b. Dependent Variable: C EMRI

The adjusted R square is 0.732 i.e. 73.2% of CEMRI is explained by the independent variables i.e. the model is fitted.

**ANOVA[a]**

| Model | | Sum of Squares | df | Mean Square | F | Sig. |
|---|---|---|---|---|---|---|
| 1 | Regression | 245.546 | 3 | 81.849 | 4.636 | .326[b] |
| | Residual | 17.654 | 1 | 17.654 | | |
| | Total | 263.200 | 4 | | | |

a. Dependent Variable: CEMRI

b. Predictors: (Constant), SP, TS, BS

The ANOVA table showed that there is a significant relationship between CEMRI and the independent variables.

**Coefficients[a]**

| Model | | Unstandardized Coefficients | | Standardized Coefficients | t | Sig. | 95.0% Confidence Interval for B | |
|---|---|---|---|---|---|---|---|---|
| | | B | Std. Error | Beta | | | Lower Bound | Upper Bound |
| 1 | (Constant) | -56.354 | 34.358 | | -1.640 | .349 | -492.914 | 380.207 |
| | TS | .001 | .001 | .529 | 2.001 | .295 | -.006 | .008 |
| | BS | 5.028 | 2.110 | .759 | 2.383 | .253 | -21.777 | 31.833 |
| | SP | -.083 | .331 | -.079 | -.251 | .843 | -4.284 | 4.118 |

a. Dependent Variable:C EMRI

The table of coefficients showed that all the independent variables are significantly repressor of CEMRI.

Therefor multiple regression model of textile industry is :

CEMRI:  -56.354+ .001Ts+5.028BS- 0.83SP.

**4.1.2 Pharmaceutical:** There are 28 listed pharmaceutical companies in DSE. We have selected 15 conveniently. The descriptive statistics is given below :

**Descriptive Statistics of Pharmaceuticals**

|  | Range | Minimum | Maximum | Mean | | Std. Deviation | Variance |
|---|---|---|---|---|---|---|---|
|  | Statistic | Statistic | Statistic | Statistic | Std. Error | Statistic | Statistic |
| BS | 2.00 | 9.00 | 11.00 | 10.2000 | .48990 | 1.09545 | 1.200 |
| CEMRI | 24.00 | 35.00 | 59.00 | 45.6000 | 4.09390 | 9.15423 | 83.800 |
| SP | 420.00 | 141.00 | 561.00 | 293.6000 | 79.55979 | 177.90110 | 31648.800 |
| TA | 12500.00 | 12000.00 | 24500.00 | 17100.0000 | 2135.41565 | 4774.93455 | 22800000.000 |
| TS | 10500.00 | 7500.00 | 18000.00 | 12540.0000 | 1742.58429 | 3896.53692 | 15183000.000 |
| Valid N (listwise) |  |  |  |  |  |  |  |

The regression model  is in the following table

**Model Summary[b]**

| Model | R | R Square | Adjusted R Square | Std. Error of the Estimate | Durbin-Watson |
|---|---|---|---|---|---|
| 1 | 1.000[a] | 1.000 | . | . | 2.500 |

a. Predictors: (Constant), SP, TS, TA, BS

b. Dependent Variable: CEMRI

The value of R-square is 1. That means the explanatory variables are not capable in explaining the dependent variable. Therefore our model is not fit in case of pharmaceuticals.

**ANOVAᵃ**

| Model | | Sum of Squares | df | Mean Square | F | Sig. |
|---|---|---|---|---|---|---|
| 1 | Regression | 114.000 | 4 | 28.500 | . | .ᵇ |
| | Residual | .000 | 0 | . | | |
| | Total | 114.000 | 4 | | | |

a. Dependent Variable: CEMRI

b. Predictors: (Constant), SP, TS, TA, BS

The coefficients and intercept is shown below.

**Coefficientsᵃ**

| Model | | Unstandardized Coefficients | | Standardized Coefficients | t | Sig. | 95.0% Confidence Interval for B | |
|---|---|---|---|---|---|---|---|---|
| | | B | Std. Error | Beta | | | Lower Bound | Upper Bound |
| 1 | (Constant) | 96.858 | .000 | | . | . | 96.858 | 96.858 |
| | TS | -.002 | .000 | -1.319 | . | . | -.002 | -.002 |
| | TA | .002 | .000 | 1.843 | . | . | .002 | .002 |
| | BS | -4.952 | .000 | -1.016 | . | . | -4.952 | -4.952 |
| | SP | -.007 | .000 | -.219 | . | . | -.007 | -.007 |

a. Dependent Variable: CEMRI

**4.1.3.Food and Allied Industries:** There are 18 food and allied industries in DSE. We selected 10 randomly as sample for the study. The descriptive statistics and regression analysis are shown in the following tables.

**Descriptive Statistics**

| | N | Minimum | Maximum | Mean | Std. Deviation |
|---|---|---|---|---|---|
| BS | 5 | 5.00 | 5.00 | 5.0000 | .00000 |
| CEMRI | 5 | 20.00 | 33.00 | 24.8000 | 4.91935 |
| SP | 5 | 18.00 | 25.00 | 21.0000 | 2.91548 |
| TA | 5 | 3378.00 | 4500.00 | 3821.6000 | 438.13560 |
| TS | 5 | 586.00 | 819.00 | 750.2000 | 93.68404 |
| Valid N (listwise) | 5 | | | | |

Model

**Model Summary[b]**

| Model | R | R Square | Adjusted R Square | Std. Error of the Estimate | Durbin-Watson |
|---|---|---|---|---|---|
| 1 | .988[a] | .977 | .907 | 1.50065 | 2.451 |

a. Predictors: (Constant), TA, SP, TS

b. Dependent Variable: CEMRI

The regression model showed that 97.77% of dependent variable i.e. CEMRI is explained by the independent variables.

**ANOVA[a]**

| Model | | Sum of Squares | df | Mean Square | F | Sig. |
|---|---|---|---|---|---|---|
| 1 | Regression | 94.548 | 3 | 31.516 | 13.995 | .193[b] |
| | Residual | 2.252 | 1 | 2.252 | | |
| | Total | 96.800 | 4 | | | |

a. Dependent Variable: CEMRI

b. Predictors: (Constant), TA, SP, TS

**Coefficients[a]**

| | | Unstandardized Coefficients | | Standardized Coefficients | | | 95.0% Confidence Interval for B | |
|---|---|---|---|---|---|---|---|---|
| Model | | B | Std. Error | Beta | t | Sig. | Lower Bound | Upper Bound |
| 1 | (Constant) | 6.104 | 27.938 | | .218 | .863 | -348.885 | 361.094 |
| | TS | -.016 | .020 | -.301 | -.790 | .574 | -.271 | .239 |
| | SP | .000 | .311 | .000 | .001 | .999 | -3.949 | 3.949 |
| | TA | .008 | .004 | .712 | 1.833 | .318 | -.047 | .063 |

a. Dependent Variable: CEMRI

The regression equation is that

CEMRI:  6.14-0.16TS+0.00BS+ 0.08SP.

The equation showed that board size does not influence CEMRI.

**4.1.4. Ceramic Industries:** There are five ceramic industries in DSE out of them 3 has been selected for the study as sample. The descriptive statistics and regression model are given below:

**Descriptive Statistics f Ceramic industries**

| | N | Range | Minimum | Maximum | Mean | | Std. Deviation | Variance |
|---|---|---|---|---|---|---|---|---|
| | Statistic | Statistic | Statistic | Statistic | Statistic | Std. Error | Statistic | Statistic |
| BS | 5 | .00 | 5.00 | 5.00 | 5.0000 | .00000 | .00000 | .000 |
| CEMRI | 5 | 30.00 | 30.00 | 60.00 | 46.8000 | 5.07346 | 11.34460 | 128.700 |
| SP | 5 | 9.00 | 61.00 | 70.00 | 66.4000 | 1.60000 | 3.57771 | 12.800 |
| TA | 5 | 2057.00 | 9100.00 | 11157.00 | 9984.8000 | 387.23267 | 865.87857 | 749745.700 |
| TS | 5 | 972.00 | 4689.00 | 5661.00 | 5037.8000 | 166.70975 | 372.77433 | 138960.700 |
| Valid N (listwise) | 5 | | | | | | | |

The table showed that CEMRI ranged between 30% to 60% and on average 47%.

**Model Summary**

| Model | R | R Square | Adjusted R Square | Std. Error of the Estimate |
|---|---|---|---|---|
| 1 | .938ª | .879 | .516 | 7.88931 |

a. Predictors: (Constant), TA, SP, TS

The model is fitted as the CEMRI is explained by 88% by the independent variables.

**ANOVAª**

| Model | | Sum of Squares | df | Mean Square | F | Sig. |
|---|---|---|---|---|---|---|
| 1 | Regression | 452.559 | 3 | 150.853 | 2.424 | .434[b] |
| | Residual | 62.241 | 1 | 62.241 | | |
| | Total | 514.800 | 4 | | | |

a. Dependent Variable: CEMRI

b. Predictors: (Constant), TA, SP, TS

**Coefficients[a]**

| Model | Unstandardized Coefficients | | Standardized Coefficients | | | 95.0% Confidence Interval for B | |
|---|---|---|---|---|---|---|---|
| | B | Std. Error | Beta | t | Sig. | Lower Bound | Upper Bound |
| (Constant) | -164.517 | 179.473 | | -.917 | .528 | -2444.932 | 2115.897 |
| TS | -.009 | .040 | -.280 | -.215 | .865 | -.513 | .496 |
| SP | 1.068 | 2.152 | .337 | .496 | .707 | -26.275 | 28.412 |
| TA | .018 | .021 | 1.401 | .876 | .542 | -.248 | .285 |

a. Dependent Variable: CEMRI

The regression model is CEMR = -164.517 - .009TS + 1.068SP+ 018TA.

**4.1.5. Fuel and Power**: There are 18 fuel and power industry listed in DSE. Conveniently 10 out of 18 have been selected for the study. The descriptive statistics of the sampled firms and regression model are given below:

**Descriptive Statistics of Fuel and Power**

| | N | Minimum | Maximum | Mean | Std. Deviation | Variance |
|---|---|---|---|---|---|---|
| BS | 5 | 7.00 | 7.00 | 7.0000 | .00000 | .000 |
| CEMRI | 5 | 20.00 | 35.00 | 27.6000 | 6.02495 | 36.300 |
| SP | 5 | 178.00 | 234.00 | 206.2000 | 22.74203 | 517.200 |
| TA | 5 | 7650.00 | 8790.00 | 8230.8000 | 505.18878 | 255215.700 |
| TS | 5 | 987.00 | 1234.00 | 1099.6000 | 94.49762 | 8929.800 |
| Valid N (listwise) | 5 | | | | | |

The range of CEMRI 20 to 35 and the mean is 27.6. The regression model is given below

**Model Summary**

| Model | R | R Square | Adjusted R Square | Std. Error of the Estimate |
|---|---|---|---|---|

| 1 | .988ᵃ | .977 | .907 | 1.83434 |
|---|---|---|---|---|

a. Predictors: (Constant), TA, SP, TS

The dependent variable is explained 98% by the explanatory variables. Therefore the model is highly fitted.

**ANOVAᵃ**

| Model | | Sum of Squares | df | Mean Square | F | Sig. |
|---|---|---|---|---|---|---|
| 1 | Regression | 141.835 | 3 | 47.278 | 14.051 | .193ᵇ |
| | Residual | 3.365 | 1 | 3.365 | | |
| | Total | 145.200 | 4 | | | |

a. Dependent Variable: CEMRI

b. Predictors: (Constant), TA, SP, TS

**Coefficientsᵃ**

| Model | | Unstandardized Coefficients | | Standardized Coefficients | t | Sig. | 95.0% Confidence Interval for B | |
|---|---|---|---|---|---|---|---|---|
| | | B | Std. Error | Beta | | | Lower Bound | Upper Bound |
| 1 | (Constant) | -73.293 | 15.565 | | -4.709 | .133 | -271.065 | 124.479 |
| | TS | -.025 | .019 | -.388 | -1.290 | .420 | -.268 | .219 |
| | SP | .040 | .045 | .151 | .880 | .540 | -.536 | .616 |
| | TA | .015 | .004 | 1.221 | 3.825 | .163 | -.034 | .063 |

a. Dependent Variable: CEMRI

The regression model is CEMR = -73.293 - .025TS + 0.40SP+ 015TA. All the coefficients are significant. The Board size has no correlation with CEMRI . So this is excluded from model.

**4.1.6. Jute:** There are listed three Jute Industries in DSE and 2 out of them have been selected for the study. The descriptive statistics and regression model is given below:

**Descriptive Statistics of Jute Industries**

|  | N | Range | Minimum | Maximum | Mean | Std. Deviation | Variance |
|---|---|---|---|---|---|---|---|
| BS | 5 | .00 | 6.00 | 6.00 | 6.0000 | .00000 | .000 |
| TS | 5 | 2290.00 | 6500.00 | 8790.00 | 7684.0000 | 813.28347 | 661430.000 |
| TA | 5 | 1212.00 | 4467.00 | 5679.00 | 5016.0000 | 514.19063 | 264392.000 |
| SP | 5 | 25.00 | 40.00 | 65.00 | 50.2000 | 9.67988 | 93.700 |
| CEMRI | 5 | 18.00 | 27.00 | 45.00 | 34.6000 | 6.84105 | 46.800 |
| Valid N (listwise) | 5 |  |  |  |  |  |  |

The CEMRI ranges between 27 and 45 and on average 34.

**Model Summary[b]**

| Model | R | R Square | Adjusted R Square | Std. Error of the Estimate | Durbin-Watson |
|---|---|---|---|---|---|
| 1 | .955[a] | .912 | .649 | 4.05144 | 2.719 |

a. Predictors: (Constant), SP, TS, TA

b. Dependent Variable: CEMRI

The dependent variable is explained 91% by the explanatory variables. Therefore the model is highly fitted.

**ANOVA[a]**

| Model | | Sum of Squares | df | Mean Square | F | Sig. |
|---|---|---|---|---|---|---|
| 1 | Regression | 170.786 | 3 | 56.929 | 3.468 | .371[b] |
|  | Residual | 16.414 | 1 | 16.414 |  |  |
|  | Total | 187.200 | 4 |  |  |  |

a. Dependent Variable: CEMRI

b. Predictors: (Constant), SP, TS, TA

**Coefficients<sup>a</sup>**

| Model | Unstandardized Coefficients | | Standardized Coefficients | t | Sig. | 95.0% Confidence Interval for B | |
|---|---|---|---|---|---|---|---|
| | B | Std. Error | Beta | | | Lower Bound | Upper Bound |
| (Constant) | -26.854 | 21.667 | | -1.239 | .432 | -302.159 | 248.452 |
| TS | .000 | .003 | -.055 | -.143 | .910 | -.042 | .041 |
| TA | .014 | .006 | 1.051 | 2.470 | .245 | -.058 | .086 |
| SP | -.102 | .240 | -.145 | -.426 | .744 | -3.148 | 2.944 |

a. Dependent Variable: CEMRI

The regression model is CEMR = -26.85 - .00TS + 0.40SP+ 015TA. The Board size has no correlation with CEMRI so this has been excluded from model.

**4.6.7. Paper:** There are two paper mills listed in DSE. Out of two one is selected for the study as sample. The descriptive statistics and regression model are given below:

**Descriptive Statistics**

| | N | Range | Minimum | Maximum | Mean | Std. Deviation | Variance |
|---|---|---|---|---|---|---|---|
| BS | 5 | .00 | 5.00 | 5.00 | 5.0000 | .00000 | .000 |
| CEMRI | 5 | 5.00 | 20.00 | 25.00 | 21.6000 | 2.30217 | 5.300 |
| SP | 5 | 33.00 | 34.00 | 67.00 | 51.2000 | 12.39758 | 153.700 |
| TA | 5 | 1298.00 | 3280.00 | 4578.00 | 3796.6000 | 544.75940 | 296762.800 |
| TS | 5 | 1400.00 | 5680.00 | 7080.00 | 6295.4000 | 534.51361 | 285704.800 |
| Valid N (listwise) | 5 | | | | | | |

The range of CEMRI is 5 and varied between 20 and 25% . The  regression  model is given below

**Model Summary**

| Model | R | R Square | Adjusted R Square | Std. Error of the Estimate |
|---|---|---|---|---|

| 1 | .998ᵃ | .995 | .981 | .31628 |

a. Predictors: (Constant), SP, TS, TA

The dependent variable is explained  99% by the explanatory variables. Therefore the model is highly fitted.

**ANOVAᵃ**

| Model | | Sum of Squares | df | Mean Square | F | Sig. |
|---|---|---|---|---|---|---|
| 1 | Regression | 21.100 | 3 | 7.033 | 70.312 | .087ᵇ |
| | Residual | .100 | 1 | .100 | | |
| | Total | 21.200 | 4 | | | |

a. Dependent Variable: CEMRI

b. Predictors: (Constant), SP, TS, TA

**Coefficientsᵃ**

| Model | Unstandardized Coefficients | | Standardized Coefficients | t | Sig. | 95.0% Confidence Interval for B | |
|---|---|---|---|---|---|---|---|
| | B | Std. Error | Beta | | | Lower Bound | Upper Bound |
| (Constant) | 13.837 | 4.583 | | 3.019 | .204 | -44.392 | 72.065 |
| TS | -.001 | .001 | -.335 | -2.490 | .243 | -.009 | .006 |
| TA | .004 | .001 | .975 | 3.457 | .179 | -.011 | .019 |
| SP | .024 | .063 | .127 | .373 | .773 | -.777 | .824 |

a. Dependent Variable: CEMRI

The regression model is CEMR = 13.83 - .001TS + 0.0040TA+ 0.24SP. The Board size has no correlation with CEMRI so this has been excluded from model

**4.2. Overall Analysis**:  The overall analysis of sampled firms are given  below:

**4.2.1.The over-all Statistics of CEMRI:** The industry-wise CEMRI is given  in the table below

Industry wise descriptive statistics of CEMRI

| Industry | Minimum % | Maximum % | Average % | Standard Deviation |
|---|---|---|---|---|
| Textile | 60 | 76 | 67 | 8 |
| Pharmaceutical | 35 | 59 | 46 | 9 |
| Food and Allied | 20 | 33 | 24 | 5 |
| Ceramic | 30 | 60 | 46 | 11 |
| Fuel and Power | 20 | 35 | 27 | 6 |
| Jute | 27 | 45 | 34 | 7 |
| Paper | 20 | 25 | 21 | 2 |

Therefore the highest CEMRI is in textile industry while the lowest is in paper mills.

**4.2.2.  Regression Equations** :  The industry wise regression equation is given below

| Industry | Regression Equation | Comment |
|---|---|---|
| Textile | CEMRI= -56.354+0.001TS+5.028BS-0.83SP. | TA(Total assets) has no correlation with CEMRI) |
| Pharmaceutical | The explanatory variables  have no correlation with the dependent variable |  CEMRI is totally independent |
| Food and Allied | CEMRI=  6.14-  0.16TS +0.00BS + 0.083SP. | Total sales has no relationship with CEMRI |
| Ceramic | CEMRI= -164.517- 0.009TS+ 1.068SP-0.018TA. | Board size has no impact on CEMRI. |
| Fuel and Power | CEMRI= -73.293-0.025TS+0.40SP-0.15TA | Board size has no correlation with CEMRI. |
| Jute | CEMRI= -26.85 - .00TS + 0.40SP+ 015TA | Board size has no correlation with CEMRI |
| Paper | CEMR = 13.83 - .001TS + 0.0040TA+ 0.24SP | Board size has no correlation with CEMRI |

# CHAPTER FIVE

## SUMMARY, FINDINGS AND POLICY IMPLICATIONS

**5.1. Summary of the Study:** The aim of first  global conference  on environment of United Nations was to   protect and improve the human environment and to remedy and prevent its impact. A set of Sustainable Development Goals (SDGs) were built  upon  the Millennium Development Goals . Among the  17 SDGs, SDG6-clean water and sanitation, SDG7-sustainable energy for all, SDG8-decent work and economic development, SDG9- innovation and infrastructure and SDG12-sustainable consumption and protection  are directly related to environmental management .In the management of environment the Environmental Management Accounting is essential  for corporate or companies  because corporate sectors are the main parties of environmental humiliation as they are existed in the environment and for protecting environment, accounting is emerged  which is called Environmental Management Accounting (EMA). Environmental Reporting (ER)   is the output of the implementation of EMA. Environmental reporting practices of the industrial organizations are crucial issues for sustainable development as  many industrial  activities gradually increase environmental hazards. Many national and international organizations have been working for the development of specific conceptual and regulatory framework such as UNEP and UNCTD. The most influential and pioneering effort on environmental reporting is Global Reporting Initiative (GRI)  and ISO 14001:2015. An increasing number of countries are imposing requirements on companies for reporting environmental performance. Denmark is the first country to adopt mandatory legislation on public environmental reporting. In Netherlands, new legislation on mandatory environmental reporting has been adopted. Both Danish and Dutch regulations require reporting to the authorities and to the public. In Norway, the new Accounting Act requires that all companies include environmental information in the annual report from 1999 onwards. In Sweden, similar legislation has been adopted for mandatory environmental disclosure in annual report. In U.S.A. the companies are required to submit data on emission of specific toxic chemicals to the Environmental Protection Agency under the Toxic Release Inventory .In Canada, The Securities Commission requires public companies to report the current and future financial or operational effects on environmental protection requirements in an Annual

Information Form. Government of Bangladesh has enacted laws regarding environment viz. Bangladesh Environment Conservation Act, 1995. The legal framework for accounting and reporting in Bangladesh is primarily governed Bangladesh Accounting Standard and Bangladesh Financial Reporting Standard ,Securities and Exchange Commission Rules 1987 and the Income Tax Ordinance-1984. These laws do not prescribe any mandatory environmental accounting disclosure by the companies. EMA will help in this case and this is the motivation behind the study.

The growth of corporate sector has simultaneously accelerated severe environmental hazards in Bangladesh .Many industries do not operate effluent treatment plants (ETP). Garments, textile and dying sectors have been developed without proper attention to their environmental consequences. Others polluting sectors are - tanneries, chemical and pharmaceutical industries and ship-breaking yards. The Department of Environment (DoE) has listed 1,176 factories that cause pollution throughout the country. Industrial growth also creates a range of problems. In consequence rapid and largely unregulated industrial development, many aquatic eco-systems are now under threat and with them the livelihood systems of local people.

There has been a growth in the voluntary environmental reporting practices of the corporate organizations worldwide. Some developed countries have initiated mandatory disclosure for corporate organizations. But in Bangladesh, corporate environmental disclosure is still in its nascent stage. Bangladeshi companies have been adopting environmental reporting practices voluntarily in recent years. The managers of corporate sectors have not clear concept on EMA and also they do not know how to implement EMA. Considering these, the researchers have undertaken the present study to fill the gap of knowledge in this area.

Inconducting the study 61 firms from environment industries have been undertaken randomely. The five years annual reports (2012-2016) of the 61 firms have been analysed with content analysis method. Under the content analysis method, first an EMA check list was developed and then the content of the annual reports were examined to see whether they reports EMA or not. The binary method i.e. if an item is depicted in annual report then "1" is puted and if not then "0". Finally an Corproate Environmental Management  Reporting Index (CEMRI) has been produced by by dividing the actual  scor by  total EMA requrments. The multiple regression was used to see whether there is any impact of corporate specific characteristics  on CEMRI where CEMRI is the dependent variable and total assets, total sales, board's size and stock price are independent varaible. The study found that the environmental management accounting in the manufacturing companies is in poor level. The maximum compliance is 67% and the lowest is 20% . The TA, TS BS and SP have been considered to find out the explanatory variables. In most of the cases board size does not play significant role in the practice of EMA in the sampled firms.

**5.2. Findings of the study:** The major findings from the data analyses are as follows:

5.2.1. The textile industry shows the highest percentage (67%) of CEMRI on average with 65% variance. The regression model depicts that Total Assets (TA) has no co-efficient on CEMRI.

5.2.2. In the pharmaceutical industry CEMRI is totally independent and no independent firm's financial characteristics explained the variation of CEMRI.

5.2.3. In case of Food and Allied industry Total sales has no impact on CEMRI.

5.2.4. In case of Fuel and Power, Ceramic, Jute and paper industry Board Size does not play any role in determining CEMRI.

**5.3. Policy Implications:** The foregoing discussion and analysis showed that the practice of CEMA in Bangladesh is in poor level. The corporations do some extent CEMA in terms of reporting purpose but there are a few companies who practices CEMA really. The government of Bangladesh has promulgated environmental conservative acts but their compliance is very limited. The company act 1994 and even 1993 have totally avoided the CEMA aspects. Only the Company Act 1994 has stated about usage of power and energy by the companies. Considering these the researchers attempted to recommend the following policy implications for the betterment of the industry:

(i) The compliance of environmental acts existing in Bangladesh should be strictly monitored by the concerned implementation bodies.

(ii) The common environmental management issues should be identified and then steps should be taken to ensure its compliance.

(iii) The industry specific EMA codes should be developed.

(iii) The board members and executives   of the corporations should be aware of EMA codes through proper training and workshops.

(iv) In approving the new industry, EMA should be ensured.

(v) The monitoring agencies should confirm the EMA in the corporate sectors..